\documentclass[pdflatex,sn-basic]{sn-jnl} 

\jyear{2021}%
\theoremstyle{thmstyleone}%
\theoremstyle{thmstyletwo}%
\theoremstyle{thmstylethree}%
\usepackage{xcolor}
\usepackage{amsmath, amssymb,bookmark}
\usepackage[hyperref]{backref}
\let\oldbibliography\thebibliography
\renewcommand{\thebibliography}[1]{%
  \oldbibliography{#1}%
  \setlength{\itemsep}{1.95pt}%
}
\newcommand\mycolor[1]{\textcolor{black}{#1}}

\begin{document}

\title[Particle acceleration by magnetic reconnection in geospace]{Particle acceleration by magnetic reconnection in geospace}



\author*[1]{\fnm{Mitsuo} \sur{Oka}}\email{moka@berkeley.edu}
\author[2,3]{\fnm{Joachim} \sur{Birn}}
\author[4]{\fnm{Jan} \sur{Egedal}}
\author[3]{\fnm{Fan} \sur{Guo}}
\author[5,6]{\fnm{Robert E.} \sur{Ergun}}
\author[7]{\fnm{Drew L.} \sur{Turner}}
\author[8]{\fnm{Yuri} \sur{Khotyaintsev}}
\author[9]{\fnm{Kyoung-Joo} \sur{Hwang}}
\author[7]{\fnm{Ian J.} \sur{Cohen}}
\author[10]{\fnm{James F.} \sur{Drake}}
\affil*[1]{\orgdiv{Space Sciences Laboratory}, \orgname{University of California Berkeley}, \orgaddress{\street{7 Gauss Way}, \city{Berkeley}, \postcode{94720}, \state{CA}, \country{USA}}}
\affil[2]{\orgdiv{Center for Space Plasma Physics}, \orgname{Space Science Institute}, \orgaddress{\street{4765 Walnut Street}, \city{Boulder}, \postcode{80301}, \state{CO}, \country{USA}}}
\affil[3]{\orgname{Los Alamos National Laboratory}, \orgaddress{\city{Los Alamos}, \postcode{87545}, \state{NM}, \country{USA}}}
\affil[4]{\orgdiv{Department of Physics}, \orgname{University of Wisconsin-Madison}, \orgaddress{\street{1150 University Avenue}, \city{Madison}, \postcode{53706}, \state{WI}, \country{USA}}}
\affil[5]{\orgdiv{Laboratory for Atmospheric and Space Physics}, \orgname{University of Colorado}, \orgaddress{\street{1234 Innovation Drive}, \city{Boulder}, \postcode{80303}, \state{CO}, \country{USA}}}
\affil[6]{\orgdiv{Department of Astrophysical and Planetary Sciences}, \orgname{University of Colorado}, \orgaddress{\street{2000 Colorado Avenue}, \city{Boulder}, \postcode{80309}, \state{CO}, \country{USA}}}
\affil[7]{\orgname{The Johns Hopkins Applied Physics Laboratory}, \orgaddress{\street{11100 Johns Hopkins Road}, \city{Laurel}, \postcode{20723}, \state{MD}, \country{USA}}}
\affil[8]{\orgname{Swedish Institute of Space Physics}, \orgaddress{\city{Uppsala}, \postcode{75121}, \country{Sweden}}}
\affil[9]{\orgname{Southwest Research Institute}, \orgaddress{\street{6220 Culebra Road}, \city{San Antonio}, \postcode{78238}, \state{TX}, \country{USA}}}
\affil[10]{\orgdiv{Department of Physics, the Institute for Physical Science and Technology and the Joint Space Science Institute}, \orgname{University of Maryland}, \city{College Park}, \postcode{20742}, \state{MD}, \country{USA}}


\abstract{Particles are accelerated to very high, non-thermal energies during explosive energy-release phenomena in space, solar, and astrophysical plasma environments. While it has been established that magnetic reconnection plays an important role in the dynamics of Earth’s magnetosphere, it remains unclear how magnetic reconnection can further explain particle acceleration to non-thermal energies. Here we review recent progress in our understanding of particle acceleration by magnetic reconnection in Earth’s magnetosphere. With improved resolutions, recent spacecraft missions have enabled detailed studies of particle acceleration at various structures such as the diffusion region, separatrix, jets, magnetic islands (flux ropes), and dipolarization front. With the guiding-center approximation of particle motion, many studies have discussed the relative importance of the parallel electric field as well as the Fermi and betatron effects. However, in order to fully understand the particle acceleration mechanism and further compare with particle acceleration in solar and astrophysical plasma environments, there is a need for further investigation of, for example, energy partition and the precise role of turbulence.}


\maketitle



\section{Introduction}\label{sec:intro}


\subsection{Motivation and Structure}
\label{sec:background}

Particles are accelerated to very high, non-thermal energies during explosive energy-release phenomena in space, solar, and astrophysical plasma environments. Unlike remote-sensing measurements of distant astrophysical objects that are often difficult to resolve spatially, {\it in-situ} measurements of Earth's magnetosphere provide unique opportunities to directly study particle acceleration and its spatial and temporal variations down to the kinetic scale. In fact, through decades of study, it is now established that  magnetic reconnection — a plasma process that converts magnetic energy into particle energy — plays an important role in the dynamics of the energy-release process in the magnetotail \mycolor{ \citep[e.g.][and references therein]{ZweibelE_2009, JiHantao_2011, HwangKJ_2023, FuselierS_2023}}. However, it remains unclear how magnetic reconnection can further explain particle acceleration to non-thermal energies (typically $\gtrsim$ 10 keV) during explosive energy-release phenomena in Earth's magnetosphere, although significant progress has been made in the past decades with spacecraft missions such as {\it Geotail, WIND, Cluster, THEMIS/ARTEMIS}, and {\it MMS}, combined with theories and simulations.

Thus, the main purpose of this paper is to review the most recent advances in our understanding of particle acceleration by magnetic reconnection in geospace which includes the magnetotail and the dayside magnetosphere. Many observational reports of particle acceleration come from the magnetotail probably because the environmental parameter $m_iV_A^2$ can be much larger in the magnetotail, where $m_i$ is the ion mass and $V_A$ is the Alfv\'{e}n speed and therefore the energization, both heating and acceleration to non-thermal energies, becomes significant  \citep[e.g.][]{PhanT_2013, ShayM_2014, Oka_2022}. 

It should be noted that the term `particle acceleration' typically refers to the process of energizing particles to non-thermal energies and does not include the meaning of heating, an increase of the plasma temperature. Therefore, a discussion of particle acceleration usually involves a power-law form of energy spectrum. However, in some cases, the term `acceleration' is used in its literal sense, as shown in the equation of motion, $ma=F$ where $m, a,$ and $F$ represent the particle mass, acceleration, and force, respectively. This usage does not differentiate between thermal and non-thermal components.  For example, Fermi acceleration in the guiding-center approximation (which will be discussed in the following section) applies to both thermal and non-thermal particles. In this paper, we have attempted to use the phrase `acceleration to non-thermal energies' when the discussion pertains to the non-thermal component. Also, we sometimes used the term `energization' when we do not differentiate thermal and non-thermal components.

\begin{figure}[b]
\centering
\includegraphics[width=1.0\textwidth]{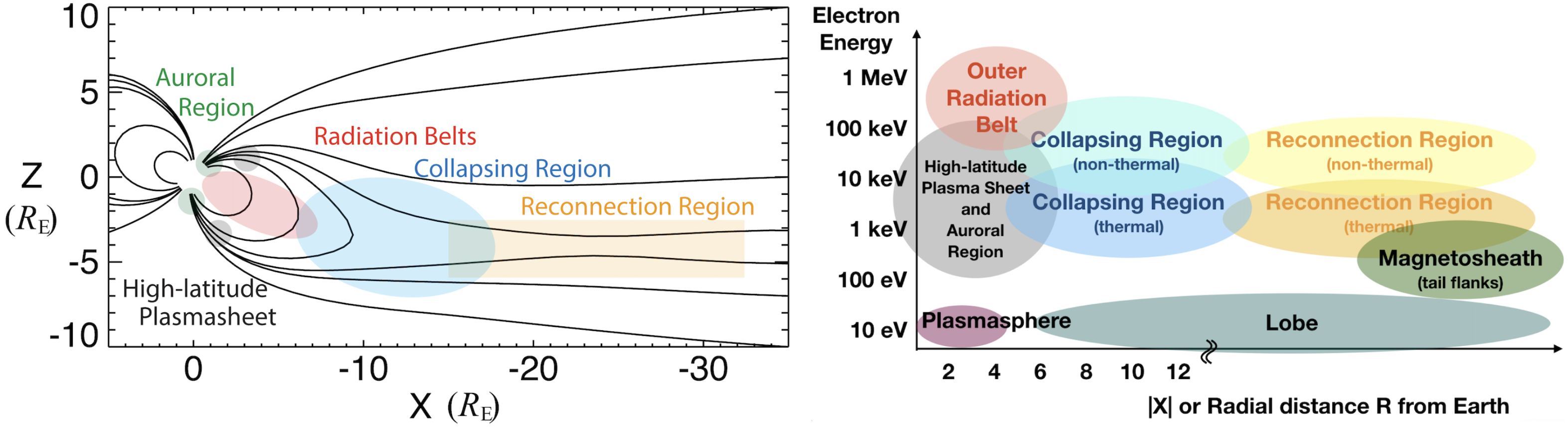}
\caption{Schematic illustrations of Earth's magnetotail, demonstrating key regions and the typical electron energy in those regions. Left: Magnetic field lines of a semi-empirical model, with the key regions highlighted in color. The Geocentric Solar Ecliptic (GSE) coordinate is used with the unit of Earth’s radii $R_E \sim 6371$ km. Right: The typical electron energy of key regions as a function of distance from Earth. Adapted from \cite{OkaM_2018}.}
\label{fig:regions}
\end{figure}

There are already relevant review articles on particle acceleration in geospace that focus on theories \citep[e.g.][]{BirnJ_2012, LiXiaocan_2021} and specific topics such as power-law index \citep{OkaM_2018} and dipolarization front \citep{FuHS_2020}. However, this paper will provide a more general overview of observations and simulations of particle acceleration to non-thermal energies both near the `reconnection region' (highlighted in yellow in Fig.  \ref{fig:regions}), which is referred to as X-line in simplified (e.g., two-dimensional or north-south symmetric) geometry, and at large scale where the intrinsic dipole field of the magnetosphere becomes important (i.e., the `collapsing region' as highlighted in blue in Fig.  \ref{fig:regions}). \mycolor{For an up-to-date overview of the relevant context of magnetic reconnection at global scales and its associated cross-scale aspects, readers are referred to \cite{FuselierS_2023} and \cite{HwangKJ_2023} in this collection, respectively.} 

The paper is structured as follows: 
\begin{enumerate}
    \item[] {\bf \ref{sec:intro}. Introduction}
    \begin{itemize}
        \item[] \ref{sec:background}: Motivation and structure
        \item[] \ref{sec:theories}: Key theories
        \item[] \ref{sec:examples}: Example observations and challenges
    \end{itemize}
    \item[] {\bf \ref{sec:X-line}. Particle acceleration near the X-line}
    \begin{itemize}
        \item[] \ref{sec:active}: Active vs quiet
        \item[] \ref{sec:fermi}: Fermi vs betatron
        \item[] \ref{sec:EDR}: Parallel electric field
        \item[] \ref{sec:turbulence}: Waves and turbulence
    \end{itemize}
    \item[] {\bf \ref{sec:largescale}. Particle acceleration at large scales}
    \begin{itemize}
        \item[] \ref{sec:largescaleoverview}: Overview
        \item[] \ref{sec:anisotropies}: Anisotropies in dipolarization events
        \item[] \ref{sec:mechanisms}: Acceleration mechanisms
        \item[] \ref{sec:sources}: Sources and seeding
        \item[] \ref{sec:diamagnetic}: Diamagnetic cavities
    \end{itemize}
    \item[] {\bf \ref{sec:discussion}. Outstanding problems}
    \begin{itemize}
        \item[] \ref{sec:partition}: Energy partition
        \item[] \ref{sec:precise}: Precise role of turbulence
    \end{itemize}
    \item[] {\bf \ref{sec:summary}. Summary and conclusion}
\end{enumerate}

\subsection{Key theories}
\label{sec:theories}

\subsubsection{Particle acceleration mechanisms in the guiding-center approximation}
\label{sec:GCA}

Observations \citep[e.g.][]{BaumjohannW_2007} and simulations have shown that the thickness of a current sheet ought to be less than a typical ion gyroradius or inertial length to enable reconnection or other activity. Thus, ions are expected to be non-adiabatic near the reconnection sites. In contrast, electrons may show adiabatic behavior much closer to an X-line, such that a guiding center approach seems more reasonable. 

\begin{figure}[t]%
\centering
\includegraphics[width=\textwidth]{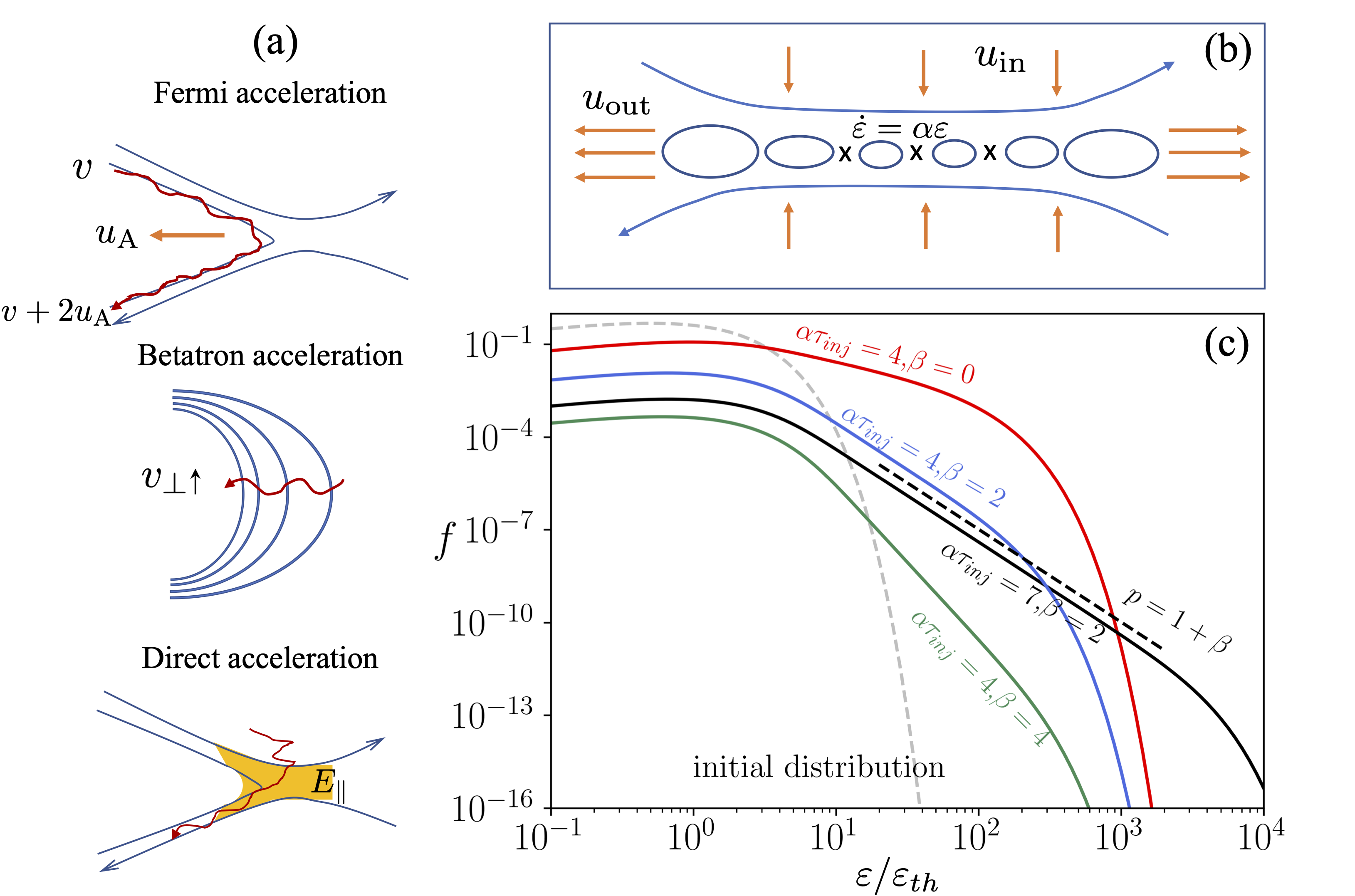}
\caption{a) Illustration of the main particle acceleration patterns, b) a cartoon for particle acceleration in the reconnection layer and development of power-law distribution, c) solutions to Equation \ref{Eq-energy-continuity} that show the energy spectra for a few different $\alpha \tau_{inj}$ and escape parameter $\beta$.}
\label{fig:theory}
\end{figure}

In the guiding-center approximation, the main acceleration mechanisms are Fermi acceleration, betatron acceleration, and the direct acceleration by the parallel electric field \citep[e.g.][]{NorthropTG_1963, BirnJ_2012, DahlinJT_2020, LiXiaocan_2021}. Fermi acceleration occurs when a particle encounters a dynamically evolving, curved magnetic field. The betatron acceleration describes the process where the increasing magnetic field leads to the energy gain in the perpendicular direction due to the conservation of the first adiabatic invariant, whereas during the direct acceleration particles stream along the magnetic field and gain energy if a significant parallel electric field exists.  Fig.  \ref{fig:theory}a shows several patterns where these acceleration mechanisms may happen. The main energy gain of a single particle under the guiding-center limit is: 
\begin{equation}
    \frac{d\varepsilon}{dt} = q \textbf{E}_\parallel \cdot  \textbf{v}_\parallel + \frac{\mu}{\gamma}\left(\frac{\partial B}{\partial t} + \mathbf{u_E}\cdot \nabla B\right) + \gamma m_e v^2_\parallel (\mathbf{u_E} \cdot \boldsymbol{\kappa})
\label{eq:energy_gain}
\end{equation}
Here, \mycolor{$q$ is the particle charge, $m_e$ is the electron mass, $\mathbf{v_{\parallel}}$ and $\mathbf{v_{\perp}}$ are the parallel and perpendicular component of the particle velocity,} $\mu$ is the magnetic moment, $\gamma$ is the Lorentz factor, and $\mathbf{u_E} = \textbf{E}\times \textbf{B}/B^2$ is the electric drift velocity, $\boldsymbol{\kappa}$ is the curvature of magnetic field lines.
The first term on the right is the parallel electric field acceleration, the second term corresponds to the betatron acceleration, and the third term is associated with the Fermi acceleration, corresponding to the curvature drift acceleration. \mycolor{In Fig. \ref{fig:theory}a, the Fermi acceleration is assumed to be driven by the curved magnetic field that drifts at the Alfv\'{e}n speed $u_A$.} 

It is to be noted that Fermi acceleration is traditionally viewed as bouncing between converging magnetic mirror points (1st order Fermi acceleration of type A) \citep{NorthropTG_1963} when the energization can be inferred from conservation of the second adiabatic invariant. However, parallel, i.e., Fermi acceleration may also result from an encounter with a strongly curved, moving magnetic field structure, akin to a slingshot effect (1st order Fermi acceleration of type B) \citep{NorthropTG_1963}. Thus, Fermi `reflections' may occur when a particle encounters a sudden change in magnetic topology and/or field strength, resulting in a sudden change in pitch angle and potentially a reflection. If the sudden change in topology is associated with a moving magnetic structure, then the particle gains energy \mycolor{corresponding} to the speed of the structure in the particle’s rest frame (i.e., \mycolor{the rate of energy gain is nearly proportional to the particle’s energy}). For electrons, a single encounter produces only a relatively small energy gain, because the speed of the structure is small compared to the electron thermal speed. However, multiple encounters can add up to substantial energy gains. Such Fermi reflections can occur in collapsing closed field regions (Fig. \ref{fig:regions}) \citep{BirnJ_2004} or in contracting or merging islands \citep{DrakeJ_2006, ZankGP_2014}.  

It is also useful to note that betatron and first-order Fermi acceleration can be viewed as $\mathbf{E}\times\mathbf{B}$ drift toward increasing magnetic field strength or in the direction of a magnetic field curvature vector, respectively (Eq. (\ref{eq:energy_gain})), but equivalently also as grad B drift or curvature drift, respectively, in the direction of an electric field (opposite for electrons) \citep[e.g.][]{BirnJ_2013}. The former indicates that the relative role of the two mechanisms depend on the magnetic field geometry, while the latter indicates a pitch angle dependence \mycolor{(as the curvature drift speed depends on the parallel particle energy and grad B drift depends on the perpendicular one)} . 

One can statistically evaluate the importance of the mechanisms by ensemble averaging particle motions, where the collective perpendicular particle current density for each species $s$ is:
\begin{equation}
\textbf{J}_{s\perp} = p_{s\parallel} \frac{\textbf{B}\times (\textbf{B} \cdot \nabla) \textbf{B}}{B^4} + p_{s\perp} \frac{\textbf{B} \times \nabla \textbf{B}}{B^3} - [\nabla \times \frac{p_{s\perp}\textbf{B}}{B^2} ] + \rho_s\textbf{u}_E - n_sm_s \frac{d\textbf{v}_s}{dt} \times \frac{\textbf{B}}{B^2} 
\label{Eq-current}
\end{equation}
\mycolor{where $p_{s\parallel}$ and $p_{s\perp}$ are parallel and perpendicular pressures to the local magnetic field, respectively, $\rho_s$ is the charge density, $n_s$ is particle number density, $m_s$ is particle mass, $\mathbf{v_s}$ is the species flow velocity, and $d/dt \equiv \partial_t + \mathbf{v_s} \cdot \nabla$.} The terms on the right shows the current due to curvature drift, grad B drift, the perpendicular magnetization, electric drift, and the polarization drift. The total energization can be shown with $\textbf{J} \cdot \textbf{E}$. Another equivalent expression for the $\textbf{J} \cdot \textbf{E}$, after a rearrangement, is
\begin{equation}
\textbf{J}_{s\perp} \cdot \textbf{E}_\perp = \nabla \cdot (p_{s\perp} \textbf{u}_E) - p_s\nabla \cdot \textbf{u}_E - (p_{s \parallel} - p_{s \perp}) \textbf{b}_i\textbf{b}_j \sigma_{ij} 
\end{equation}
\noindent \mycolor{where $\sigma_{ij} = 0.5 (\partial_i \textbf{u}_{Ej} + \partial_j \textbf{u}_{Ei} - (2\nabla \cdot \textbf{u}_E \delta_{ij}/3))$ is the shear tensor of $\mathbf{u_E}$ flow, $p_s \equiv (p_{s\parallel} + 2p_{s\perp})/3$ is the effective scalar pressure, and} we have ignored the effect of the polarization drift \citep{LiXiaocan_2017}. This expression shows the role of fluid compression and velocity shear in the energy gain \citep{LiXiaocan_2018_roles}. These can be connected with the recent work of pressure-strain terms for gaining insight in turbulent plasmas \citep[e.g.,][]{YangYan_2017,DuSenbei_2018,LiXiaocan_2019_massratio}. 

Over the past decade, particle-in-cell simulations and test-particle simulations have been widely used to evaluate these acceleration mechanisms. Several particle kinetic simulations  \citep[e.g.][]{GuoFan_2014,DahlinJ_2014,LiXiaocan_2017, ArnoldH_2021}, modeling the formation of multiple magnetic islands or flux ropes and their merging, indicated an overall dominance of Fermi acceleration over betatron acceleration. In contrast, a test-particle simulation of electron drifts in a collapsing magnetic arcade with strong guide field  indicated a dominance of betatron acceleration \citep{BirnJ_2017_Solar}. This confirms that the relative role of the two mechanisms depends on the field geometry, for instance, betatron acceleration may be expected to be important particularly inside of reconnection jet fronts and collapsing magnetic traps. In addition, parallel electric fields are shown to contribute to particle energization and modify the distribution functions \citep[e.g.][]{EgedalJ_2009}. As the guide field increases, the Fermi acceleration becomes less efficient, but acceleration by the parallel electric field is not very sensitive to the guide field \citep{DahlinJ_2016,LiXiaocan_2018_roles}.

\subsubsection{Formation of nonthermal power-law energy spectra in reconnection acceleration}
Power-law energy spectra are a main feature of nonthermal acceleration and are of great interest to reconnection studies \mycolor{\citep[e.g.][]{LiXiaocan_2019,ZhangQile_2021,ArnoldH_2021,NakanotaniM_2022}}. There have been debates and some confusion about the formation of nonthermal power-law energy spectra during particle acceleration in magnetic reconnection; therefore this issue is worth clarifying. As shown by Fig.  \ref{fig:theory}b, we illustrate a simple case where the main acceleration term is a Fermi-like relation $\dot{\varepsilon} = \alpha \varepsilon$ \mycolor{($\alpha$ is the acceleration rate}) in an energy continuity equation \citep{GuoFan_2014,GuoFan_2015}, which represents the case when the first-order Fermi acceleration dominates the acceleration process:

\begin{equation}
    \frac{\partial f}{\partial t} + \frac{\partial}{\partial \varepsilon} (\dot{\varepsilon} f) = \frac{f_{inj}}{\tau_{inj}} - \frac{f}{\tau_{esc}}
\label{Eq-energy-continuity}
\end{equation}

As reconnection proceeds, the ambient plasma is continuously injected into the reconnection layer through an inflow speed $u_{in}$. $\tau_{inj}$ is the timescale for the injection of low-energy particles $f_{inj}$, and $\tau_{esc}$ is the escape timescale. For illustration purposes, we assume that the upstream distribution is a Maxwellian distribution $f_{inj} = (2N_{inj}/\sqrt{\pi})\sqrt{\varepsilon_0}exp(-\varepsilon_0)$, where $\varepsilon_0 = \varepsilon/\varepsilon_{th}$ is energy normalized by the thermal energy. With these assumptions, the solution to Eq. (\ref{Eq-energy-continuity}) can be written as 

\begin{equation}
f(\varepsilon, t) = \frac{2 N_{inj}}{\sqrt{\pi}(\alpha \tau_{inj} \varepsilon_0^{1+\beta})}[\Gamma_{3/2+\beta}(\varepsilon_0 e^{-\alpha t}) - \Gamma(3/2+\beta)(\varepsilon_0)],
\end{equation}
where $\beta = 1/(\alpha \tau_{esc})$ and $\Gamma_s(x)$ is the upper incomplete Gamma function. Fig.  \ref{fig:theory}c illustrates this simple solution for a few different escape time and $\alpha \tau_{inj}$. As reconnection proceeds, new particles are injected and accelerated in the reconnection, \mycolor{and} a power-law distribution can form when $\alpha \tau_{inj}$ is large. Note that the derivation also shows that power-law distribution can still form even for the case with no escape term as shown in Fig. \ref{fig:theory} (red curve). However, if the population of particles is initially in the current sheet, it can be shown the distribution remains a Maxwellian \citep{GuoFan_2020}. 
It is often argued that some loss mechanism is needed to form a power-law distribution, but the simple analytical solution does not support it. Here the main physics for forming a power-law is due to the continuous injection and Fermi acceleration. Meanwhile, it is still important to understand the escape term, as it can strongly change the shape of the distribution. Other acceleration can, in principle, form a power-law, and the steady-state solution has the spectral index 

\begin{equation}
    p = 1 + \frac{1}{\alpha \tau_{esc}} + \frac{\partial \ln \alpha}{\partial \ln \varepsilon} .
\end{equation}

This equation includes the case where the acceleration rate has an energy dependence. \mycolor{When the product of $\alpha\tau_{esc}$ does not depend on energy and $\alpha$ has a power-law dependence on energy $\varepsilon$ over a certain energy range, $p$ is a constant across this range} (power-law energy spectra). 

\mycolor{In the context of magnetic reconnection, magnetic islands (or flux ropes in 3D) can play an important role in particle acceleration \citep[e.g.][]{DrakeJ_2006, DrakeJF_2013, OkaM_2010_coa, HoshinoM_2012, GuoFan_2014, ZankGP_2014, leRouxJA_2015, leRouxJA_2018}. Using a more formal, particle transport equation that captures the essential physics of particle acceleration in multi-island region,  the possibility of compressible flux-rope contraction and merging in a turbulent media was considered in \cite{ZankGP_2014, leRouxJA_2015, leRouxJA_2018}. It was shown theoretically that both curvature drift and betatron acceleration, due to an increasing flux-rope magnetic field strength, contribute to kinetic energy gain, and the particle acceleration is a first-order Fermi acceleration process when the particle distribution is isotropic or nearly isotropic. }

\subsubsection{Beyond guiding-center approximation}
Although numerical simulations have shown that the guiding-center approximation can well describe the acceleration of particles in the reconnection region, the particle dynamics in the reconnection region can be more complicated. Close to the X-line, it is well known that the gyrotropic approximation is not valid. Although the X-line may not be the region with the strongest acceleration, they may support local acceleration that is of interest to {\it in situ} observations. The X-line region with a weak guide field can support chaotic orbits of particles \citep{ZenitaniS_2016}. During particle motion, waves and turbulence can modify the particle distribution. However, as long as the particle distribution is gyrotropic, Eq. (\ref{Eq-current}) can still statistically describe the acceleration, even if significant pitch-angle scattering occurs \citep{HazeltineRD_2003, EgedalJ_2013}. Therefore, some caution is needed when interpreting the results of the analysis. In addition, waves and turbulence may lead to stochastic heating and acceleration \citep[e.g.][]{ZankGP_2015, ErgunRE_2020_sim}. The guiding-center description does not distinguish electrons and ions, meaning multiple species can be accelerated \citep[e.g.][]{ZhangQile_2021, ZhangQile_2022_arxiv}. However, in the context of the magnetosphere, ions may not be well \mycolor{described by the guiding center approximation}, as their gyroradii can be fairly large compared to characteristic scales at which the fields evolve. In particular, close to the center of the magnetotail current sheet, the gyroradii can approach the curvature radius \mycolor{of the} magnetic field, and the ions experience strong scattering when crossing the current sheet \citep[e.g.][]{RichardL_2022}. Furthermore, instabilities, waves, and turbulence that are generated during magnetic reconnection may lead to more efficient acceleration of particles \citep[e.g.][]{DahlinJ_2017, LiXiaocan_2019, ZhangQile_2021, JohnsonGrant_2022}.


\subsection{Example observations and challenges}
\label{sec:examples}

\begin{figure}[t]
\centering
\includegraphics[width=12cm]{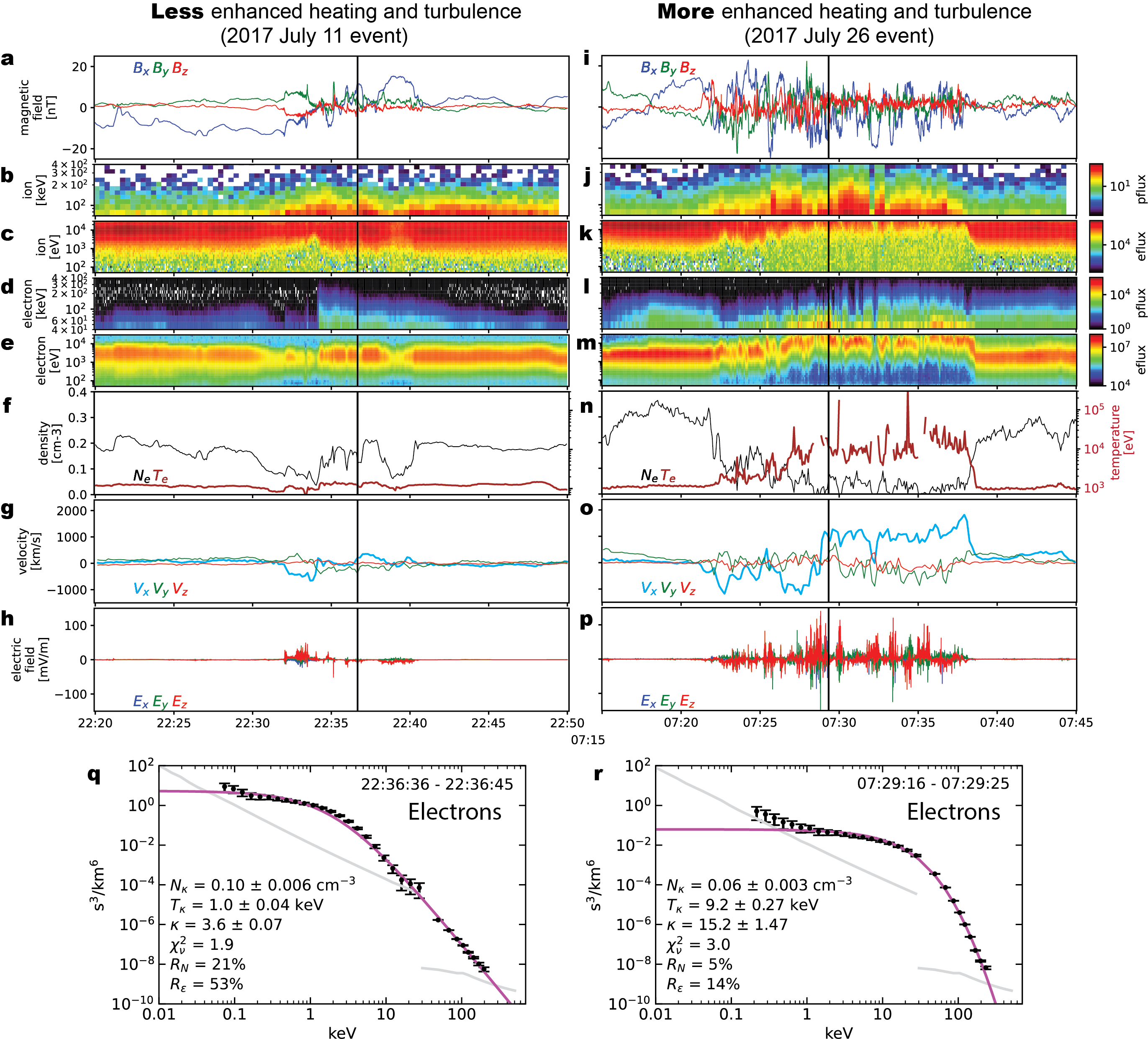}
\caption{Example observations demonstrating particle acceleration during magnetic reconnection with less enhanced and much enhanced heating/turbulence obtained on 2017 July 11 \citep[e.g.][]{TorbertRB_2018} and 2017 July 26 \citep[e.g.][]{ErgunRE_2018}, respectively. The bottom panels show the typical energy spectrum of electrons for each event obtained at the time indicated by the vertical dashed lines in the upper panels \citep{Oka_2022}. The best-fit kappa distribution model is displayed in magenta and the obtained parameters are annotated. Note the significant heating but soft (steep) power-law tail for the case of 2017 July 26.}
\label{fig:example_events}
\end{figure}

The possible limitation of the guiding-center theory and the importance of turbulence may be glimpsed in recent examples of magnetotail reconnection. Fig.  \ref{fig:example_events} shows two cases of particle acceleration during magnetotail reconnection, obtained by {\it Magnetospheric MultiScale (MMS)}. The 2017 July 11 event (left column) is a case  with less-enhanced heating and turbulence and has been studied by many authors \citep[e.g.][]{TorbertRB_2018, GenestretiKJ_2018, NakamuraTKM_2018, EgedalJ_2019, NakamuraR_2019, JiangK_2019, SitnovM_2019, BurchJL_2019, HasegawaH_2019, TorbertRB_2020, HwangKJ_2019, CohenIJ_2021, TurnerDL_2021_gyro, Oka_2022}. On the other hand, the 2017 July 26 event (right column) is a case with significantly enhanced heating and turbulence and has also been studied intensively \citep[e.g.][]{ErgunRE_2018, ErgunRE_2020_obs, ErgunRE_2020_sim, CohenIJ_2021, Oka_2022}.

A puzzle is that, some properties of magnetic reconnection (e.g., heating and turbulence) appear differently in these two cases, and yet particles (both ions and electrons) are accelerated to non-thermal energies in both cases. For electrons, the non-thermal, power-law tail may even be softer in the significantly heated and turbulent case \citep[e.g.][]{ZhouMeng_2016, Oka_2022} but it remains unclear how the observed power-law index can be explained. For ions, the energy spectrum could be more complicated. While the bulk flow component could peak around 1 keV,  the higher-energy end of the spectrum may be influenced by the physical size of the energization region, as often argued in the shock physics \citep[e.g.][]{BlandfordR_1987}. In the Earth's magnetotail, the gyroradii of ions with energies greater than $\sim$100 keV may exceed several ion skin depths. In any case, the similarities and differences of these two cases lead to questions such as  `What is the precise condition of particle acceleration?', 'What is the precise role of turbulence?', `How particle energies are partitioned between thermal and non-thermal energies?', and ultimately `How are particles heated and accelerated to non-thermal energies?'. By reviewing recent progress in more detail in this paper, we hope to clarify what we know so far, what ideas have been discussed, and what we need to work on in the near future.

\section{Particle acceleration near the X-line}\label{sec:X-line}


\subsection{Active vs. Quiet}
\label{sec:active}

Early magnetotail studies showed that, in the plasma sheet, the energy spectra become non-thermal above $\sim$10 keV for ions and $\sim$1 keV for electrons and are often represented by the kappa distribution \citep[e.g.,][and references therein]{ChristonSP_1988,ChristonSP_1989, ChristonSP_1991}.
\begin{figure}[t]
\centering
\includegraphics[width=0.8\textwidth]{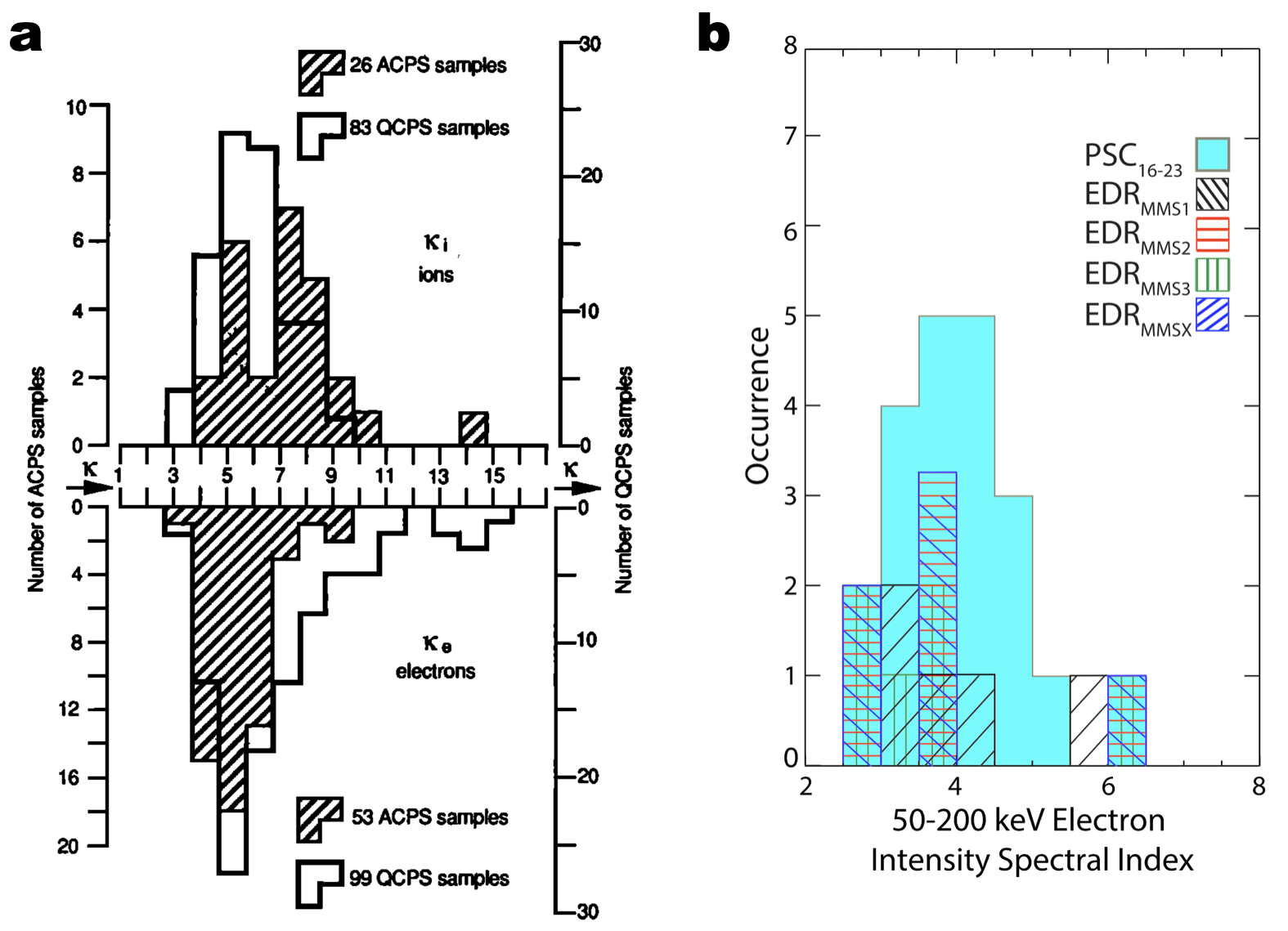}
\caption{Statistical analysis of the electron power-law index in the ISEE-era \citep[left;][]{ChristonSP_1991} and the MMS-era \citep[right;][]{CohenIJ_2021}. These analyses demonstrate that the non-thermal tail remains significant even during the quiet-time plasma sheet.}
\label{fig:christon-cohen}
\end{figure}
\mycolor{The power-law index $\kappa$ is often in the range of $\kappa \gtrsim 4$, as shown in Fig. \ref{fig:christon-cohen}a.} \cite{OierosetM_2002} showed that these energetic particles may be the result of energization occurring during reconnection in the tail. They found that the hardest particle spectrum (i.e., the most energetic particles) was observed near the center of an ion diffusion region traversed by the Wind spacecraft in the deep (60 R$_{\text{E}}$) terrestrial magnetotail. This work was followed up by \cite{CohenIJ_2021}, who explored whether a similar result would be found for a set of six electron diffusion regions discovered by MMS.
Comparing the results to a statistical dataset of 133 quiet-time (i.e., AE$^{*}$ $\lt$ 300 nT and no fast ($\mid\!v_{\text{ion avg}}\!\mid\,\ge$ 100 km/s) flows) plasma sheet crossings \mycolor{(PSC)}, the authors found that the electron diffusion region (EDR) events did in fact have harder spectra (i.e., more energetic particles) \mycolor{(Fig. \ref{fig:christon-cohen}b)}. The result suggests that these energetic electrons are coming from a local source associated with active reconnection \citep[e.g.,][]{ErgunRE_2020_obs}. In fact, an observational study reported significant heating within the EDR, followed by an appearance of the non-thermal tail in the immediate downstream of the EDR \citep[e.g.][]{OkaM_2016}. MMS observations also reported a significantly enhanced flux of energetic electrons within the EDR, although the non-thermal, power-law tail was soft with the power-law index of $\sim$8 as measured in the phase space density \citep{LiXinmin_2022}. Interestingly, \cite{TurnerDL_2021_gyro} reported coherent gyrophase bunching of $>$ 50 keV electrons in the immediate downstream of the EDR and argued that it \mycolor{can be caused by the first-order Fermi acceleration Type B off of the outflowing exhaust structure, evidencing } electron acceleration at the reconnection site and possibly also in the outflowing exhaust jets of the active reconnection.

Despite the possible importance of magnetic reconnection, it has also been reported that the non-thermal component is significant even during periods of low geomagnetic activity (AE $\lt$ 100 nT) \citep{ChristonSP_1989, CohenIJ_2021, Oka_2022}. \mycolor{This is also illustrated by the overlap of the PSC and EDR histograms in Fig. \ref{fig:christon-cohen}.} \cite{CohenIJ_2021} argued that such energetic particles may be sourced by remote down-tail reconnection sites or processes not directly related to reconnection at all. This is consistent with an earlier report of significant non-thermal tail during low geomagnetic activity \citep{ChristonSP_1989}. Similarly,  \cite{Oka_2022} examined the spatial variation across the reconnection region and reported that the non-thermal power-law tail can exist even outside the reconnection region (Hall region) where there is no significant plasma flows and turbulence. Therefore, the relationship between the production of energetic electrons and the geomagnetic activity remains unclear, let alone the importance of the EDR.


\subsection{Fermi vs betatron}
\label{sec:fermi}

As reviewed in Section \ref{sec:GCA}, the main acceleration mechanisms in the guiding-center approximation are Fermi acceleration, betatron acceleration, and the direct acceleration by the parallel electric field.  While the parallel electric field might be important for heating (as separately reviewed in Section \ref{sec:EDR}) or for acceleration to non-thermal energies in some cases \citep[e.g.][]{ZhouMeng_2016, ZhouMeng_2018}, many studies argue that Fermi and betatron acceleration are predominantly important during magnetic reconnection.

In observational studies, a pitch angle anisotropy has been the key feature for diagnosing Fermi and betatron acceleration \citep[e.g.][]{SmetsR_1999}. Particles experiencing Fermi and betatron acceleration tend to exhibit parallel and perpendicular anisotropy, respectively.  However, with the launch of MMS in 2015, electron data with the time resolution of $\sim$100 times higher than its predecessors became available. Such data sets, combined with the multi-spacecraft approach which is necessary to estimate the magnetic field curvature, have enabled us to evaluate each term in Eq. (\ref{eq:energy_gain}), providing a more direct diagnostics of the acceleration mechanism, i.e., Fermi acceleration, betatron acceleration, and the direct acceleration by the parallel electric field. Significant progress has been made with such analysis and will be reviewed below. It is to be noted that most of the discussion in this subsection is focused on electron acceleration, although there have been some studies of ion acceleration by simulations \citep[e.g.][]{BirnJ_2015_DF, UkhorskiyAY_2017} and observation \citep[e.g.][]{WangShan_2019_ions}.

\subsubsection{Outflows near the X-line}

\begin{figure}[t]
\centering
\includegraphics[width=6cm]{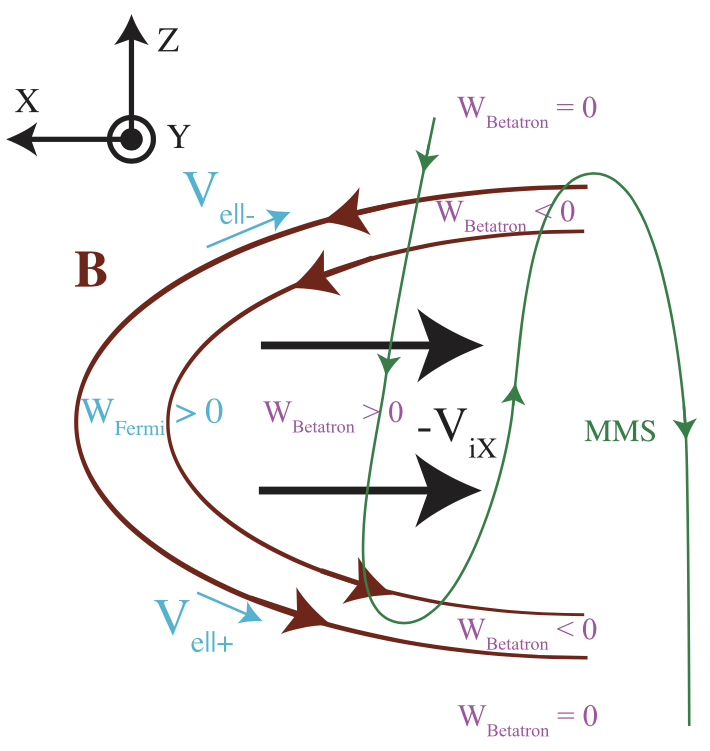}
\caption{A schematic illustration of the outflow region near the X-line, demonstrating expected Fermi and betatron acceleration signatures \citep{ErikssonE_2020}. $W_{\textrm{Fermi}}$ and $W_{\textrm{betatron}}$ represents the power-density of electron acceleration due to Fermi and betatron acceleration, respectively. } 
\label{fig:Eriksson2020}
\end{figure} 

Early studies argued that, in the outflow region immediately downstream of the X-line, magnetic field magnitude increases and that electrons are accelerated  by the gradient B and/or curvature drift \citep{HoshinoM_2001, ImadaS_2005, ImadaS_2007}. \mycolor{However, it was also argued that, above a few keV, the $\kappa$ value of electrons can approach $\sim 1$} where $\kappa^2$ is the ratio of the magnetic field curvature and the particle gyro-radius. \mycolor{In such a condition, a non-adiabatic behavior or scattering becomes important}. 

\cite{WuMingyu_2015} proposed that electrons are first pre-energized at the X-line, accelerated non-adiabatically in the pileup region in the immediate downstream region, and then further accelerated adiabatically in association with burst bulk flows (BBFs) in the outflow region. While energetic electron events tend to be rare tailward of the X-line in the tail, \cite{ChenG_2019} reported three cases of outflow jets on the tailward side and argued based on anisotropy that electrons were accelerated adiabatically by both Fermi and betatron effects. The observations were made on the tailward (or `unconfined') side where the effect of the intrinsic, dipole magnetic field can be neglected, but the outflow speeds were increasing in time (or `growing') leading to the compression or strengthening of the magnetic field.

More recently, {\it MMS} has enabled to study the acceleration mechanism with the guiding-center approximation described by Eq. (\ref{eq:energy_gain}). A case study of tailward outflows reported that the dominant mechanism, both on average and the peak values, was Fermi acceleration with a peak power density of about $+200$ pW/m$^3$ \citep{ErikssonE_2020}. During the most intense Fermi acceleration, the magnetic field curvature was comparable to the electron gyro-radius (i.e.,  $\kappa \sim 1$), suggesting electrons were being scattered efficiently. Fig.  \ref{fig:Eriksson2020} shows the schematic illustration of their interpretation. In the current sheet center, the power-density of electron acceleration due to Fermi acceleration, $W_{\textrm{Fermi}}$, and betatron acceleration, $W_{\textrm{betatron}}$, are positive because the magnetic field magnitude increases with the increasing distance from the X-line. At the edges of the current sheet, however, incoming electrons experience decreasing magnetic field and hence negative values of $W_{\textrm{betatron}}$. Interestingly, some of these findings (such as moderately non-adiabatic behaviors, energy loss at the edges, etc.) are consistent with earlier simulation results \citep{HoshinoM_2001}.

\subsubsection{Flux ropes}
A magnetic flux rope is one of the key structures associated with magnetic reconnection. It is often referred to as a magnetic island especially in 2D theoretical pictures \citep[e.g.][and references therein]{BirnJ_2012, ZankGP_2014}.  A distinction is typically made based on the absence or presence of a magnetic field component along the center of the island or rope structure. Many observations indicate that electrons are accelerated to non-thermal energies within the flux ropes both in the magnetotail \citep[e.g.][]{ChenLJ_2008, ChenLJ_2009, RetinoA_2008, WangR_2010a, WangR_2010b, HuangSY_2012, SunWeijie_2022, WangShimou_2023} and in the magnetopause \citep[e.g.][]{OierosetM_2011}. \mycolor{Also, multi-island coalescence may be a key process for the energy conversion during reconnection and associated acceleration of particles  \citep[e.g.][and references therein]{OkaM_2010_coa, leRouxJA_2015, TehWL_2023}.}

The standard theory for electron acceleration in flux ropes is the contracting island mechanism, whereby particles receive a Fermi-type energization kick at each end of an actively contracting magnetic island \citep[e.g.][]{DrakeJ_2006, ZankGP_2014, ArnoldH_2021} \cite[but see also e.g.][]{SternDP_1979, KliemB_1994}. The process requires an escape process in order to explain the observed spectral indices of energetic particles. 

Observations also indicate the importance of Fermi acceleration in addition to betatron acceleration \citep[e.g.][]{HuangSY_2012, ZhongZH_2020, JiangK_2021, SunWeijie_2022}. \cite{ZhongZH_2020} studied electron acceleration within ion-scale flux ropes by evaluating the equation for adiabatic electrons with the guiding center approximation (GCA, Eq. (\ref{eq:energy_gain}) in Section \ref{sec:GCA}). Their analysis indicated that the lower energy ($<$10 keV), field-aligned electrons experienced predominantly Fermi acceleration in a contracting flux rope, while the higher energy ($>$10 keV) electrons with perpendicular anisotropy gained energy mainly from betatron acceleration. They argued that the dominance of betatron acceleration at high energies could be a consequence of the 3D nature of the flux rope. The field-aligned electrons that can experience Fermi acceleration would quickly escape along the axis of the flux rope. Because of the successful application of the GCA theory, the study was positively commented by \cite{DahlinJT_2020}. 

\mycolor{In another case study of a pair of tailward traveling flux ropes, \cite{SunWeijie_2022} reported that, while Fermi and parallel potential is strong near the X-lines between the flux rope pair, betatron is strong on flux rope boundaries. For electron acceleration at the magnetopause, \cite{WangShimou_2023} reported an interaction of two filamentary currents (FCs) within a flux rope and argued that the electrons were mainly accelerated by the betatron mechanism in the compressed region caused by the FC interaction. }

However, electron acceleration in flux ropes might not always be adiabatic \citep[e.g.][]{OkaM_2010_surf, FujimotoK_2021, SunWeijie_2022, WangShimou_2023}, and the parallel electric fields might become important, particularly for small-scale, secondary flux ropes that form at and around the primary X-line  with intensified current \citep[e.g.][]{WangHuanyu_2017, ZhouMeng_2018, JiangK_2021}. There can also be intense wave activities, turbulence, and current filaments inside flux ropes \citep[e.g.][]{FuHS_2017, HuangSY_2019, JiangK_2021, SunWeijie_2022, WangShimou_2023} that can lead to stochastic acceleration. Recent 3D simulations have demonstrated that such turbulence and associated induced electric field can result in strong heating of electrons \citep{FujimotoK_2021}.

\subsection{Parallel electric field}
\label{sec:EDR}

\mycolor{In the earlier years of magnetic reconnection studies,} it was proposed that electrons are accelerated directly in the reconnection electric field along the magnetic X-lines \citep[e.g.][and references therein]{LitvinenkoYE_1996}. However, recent studies of magnetic reconnection have revealed a new adiabatic picture in which the parallel electric field plays an important role, \mycolor{as reviewed in this subsection.} 
\mycolor{Here, it is worth noting that, while the rate of energy gain is roughly proportional to the particle’s energy for the cases of Fermi and Betatron acceleration, the rate of energy gain scales only with the particle speed $v$ for the case of direct acceleration by parallel electric field (Section \ref{sec:theories}). Nevertheless, the acceleration by parallel electric field can boost thermal particles by orders of magnitude in energy and hereby provide a preenergized seed populations subject to further Fermi and Betatron  energization \citep{EgedalJ_2015}.}

For many plasma physics problems, it is important to understand  how rapidly  thermal (and super-thermal) electrons travel along the magnetic lines. As an example, we may consider the July 11, 2017, reconnection event recorded at about 20$R_E$ into the Earth's magnetotail with a typical electron temperature of 1 keV (See Fig. \ref{fig:example_events}, left column, in Section \ref{sec:examples}). It follows that the electron thermal speed ($v_{te}\simeq 20\cdot10^{6}$m/s) is about 400 times faster than the expected reconnection inflow speed $v_{in}\simeq v_A/10 \simeq 50\cdot10^{3}$m/s. Thus, during the course of a fluid element (say, initially  1$d_i\simeq 1\cdot10^6$m upstream of the reconnection site) traversing the reconnection region, a typical electron will travel a distance of about  \mycolor{$(v_{te}/v_{in}) d_i \simeq 80R_E$} (larger than the distance from Earth to the moon).
\mycolor{This means that electrons, once energized, would escape instantly from the energization site and would not exhibit a localized, enhanced flux at and around the energization site, if there were no confinement or trapping. In reality, however, energetic electrons are observed in the localized region of magnetic reconnection (See, for example, Fig. \ref{fig:example_events} and other studies reviewed elsewhere in this paper). Therefore, we need a model for electron confinement or trapping to explain the observations.}

\begin{figure}[t]
\centering
\includegraphics[width=12cm]{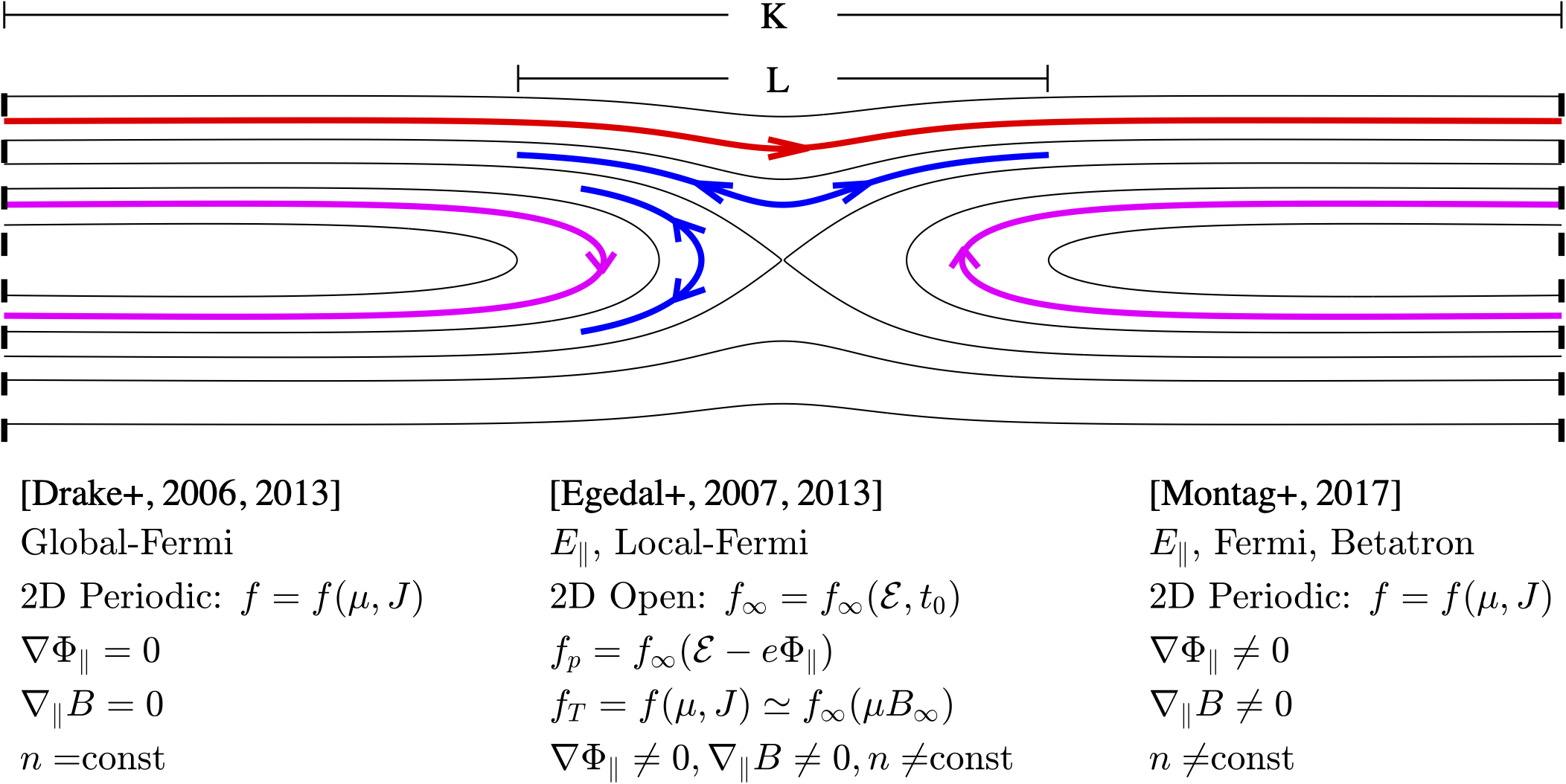}
\caption{Examples of trapped and passing electron orbits in 2D magnetic configurations, as well as key assumptions applied in adiabatic models for electron energization during magnetic reconnection. Adapted from \cite{MontagP_2017}.}
\label{fig:Egedal_orbits4}
\end{figure}

Due to the fast streaming of the electrons along the magnetic field lines, their parallel action,  $J=\oint v_{\|} dl$, is typically a well conserved adiabatic invariant. As illustrated in Fig.  \ref{fig:Egedal_orbits4}, this $J$-invariance has been explored in a range of theoretical models for electron heating. The model of \cite{DrakeJ_2006, DrakeJF_2013} considers a 2D periodic and incomprehensible system and in essence applies Jeans' theorem \citep{JeansJH_1915} that the gross evolution of the electrons is governed by a double adiabatic assumption, $f=f(J,\mu)$ where $\mu$ is the magnetic moment,  augmented with phenomenological pitch angle scattering. This  Fermi heating model is only concerned with the large-scale energization of the electrons and ignores any variation in $f$ along magnetic field lines.  

Meanwhile, \cite{EgedalJ_2008, EgedalJ_2013} assumes that the reconnection region is embedded in a large open system, where the plasma in the ambient regions provides fixed sources of electrons, $f_{\infty}$. Electrons may here be characterized as either passing or trapped. The passing electrons instantaneously travel along the field lines with their total energy conserved, $U ={\cal E}-e\Phi_{\|}$, where $\Phi_{\|}=\int_{x}^{\infty} E_{\|} dl$ is the acceleration potential \citep{EgedalJ_2009}. Meanwhile, the trapped electrons again follow Jeans' theorem, $f_T=f_{\infty}(J,\mu)$. Furthermore, with the imposed boundary conditions it can be shown that $f_T\simeq f_{\infty}(\mu B_{\infty})$, and a relatively simple analytical form is obtained:

\begin{equation}
\mycolor{f=f_{\infty}({\cal E} - e\Phi_{\|})}\,\,{\textrm{(passing)}}\,\,,\quad f= f_{\infty}(\mu B_{\infty})\,\,{\textrm{(trapped)}}\,\,.
\label{eq:felectrons}
\end{equation}

These types of  distributions are  common in measurements within reconnection regions and have been observed by multiple spacecraft missions including Wind, Cluster, THEMIS and MMS 
\citep{EgedalJ_2005, EgedalJ_2010, OkaM_2016, ErikssonE_2018, WethertonBA_2019, WethertonBA_2021, WangShimou_2021}. 
The global model by  Drake et al. and the local model in Eq. (\ref{eq:felectrons}) can be obtained as two separate limits of the more general framework recently developed by \cite{MontagP_2017}. The model by Drake et al. follows directly by imposing the conditions of $n=$ constant and $\nabla_{\|}B=0$, while  Eq. (\ref{eq:felectrons}) is recovered in the  limit $K\gg L$, \mycolor{where as illustrated in Fig. \ref{fig:Egedal_orbits4}, $K$ is the size of the periodic domain and $L$ is the typical length scale for electron trapping in the reconnection region.}

Observations suggest that the bulk electron heating for a range of reconnection scenarios is largely governed by Eq. (\ref{eq:felectrons}). The trapped electrons have negligible heat-exchange with the ambient plasma, and when the majority of the thermal electrons are trapped the pressure components along and perpendicular to the magnetic field follow the CGL \citep{ChewGF_1956} scaling laws $p_{\|} \propto n^3/B^2$ and $p_{\perp} \propto nB$. This is also the asymptotic limit (at large $n/B$) of the equation of state derived directly from Eq. (\ref{eq:felectrons}) by \cite{LeA_2009} (hereafter referred to as L\^{e}2009 EoS).

\begin{figure}[t]
\centering
\includegraphics[width=12cm]{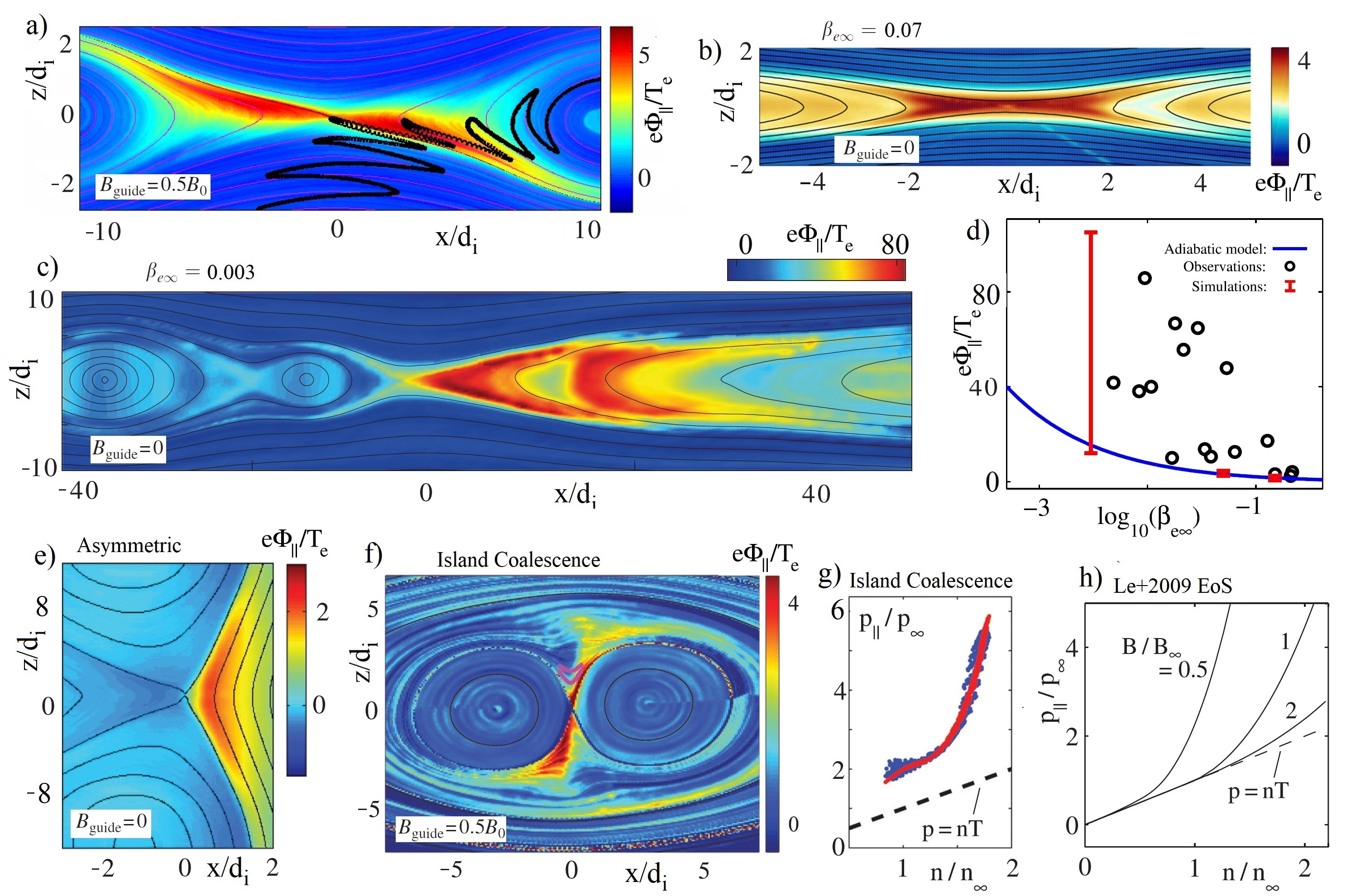}
\caption{a-c,e,f) Example profiles of $\Phi_{\|}$ observed in kinetic simulation under various reconnection scenarios. d) The blue line is the adiabatic prediction for $\Phi_{\|}$, while Cluster observations and kinetic simulations show strongly enhanced values of $\Phi_{\|}$ for low values of $\beta_{e\infty}$. g) $p_{e\|}$ as a function of $n$ observed in island coalescence demonstrating parallel heating beyond the level predicted by the L\^{e}2009 EoS (in h)). Collected and adapted from \cite{LeA_2012, LeA_2016} and \cite{EgedalJ_2013}.}
\label{fig:Egedal_Potentials}
\end{figure} 

\begin{figure}[t]
\centering
\includegraphics[width=12cm]{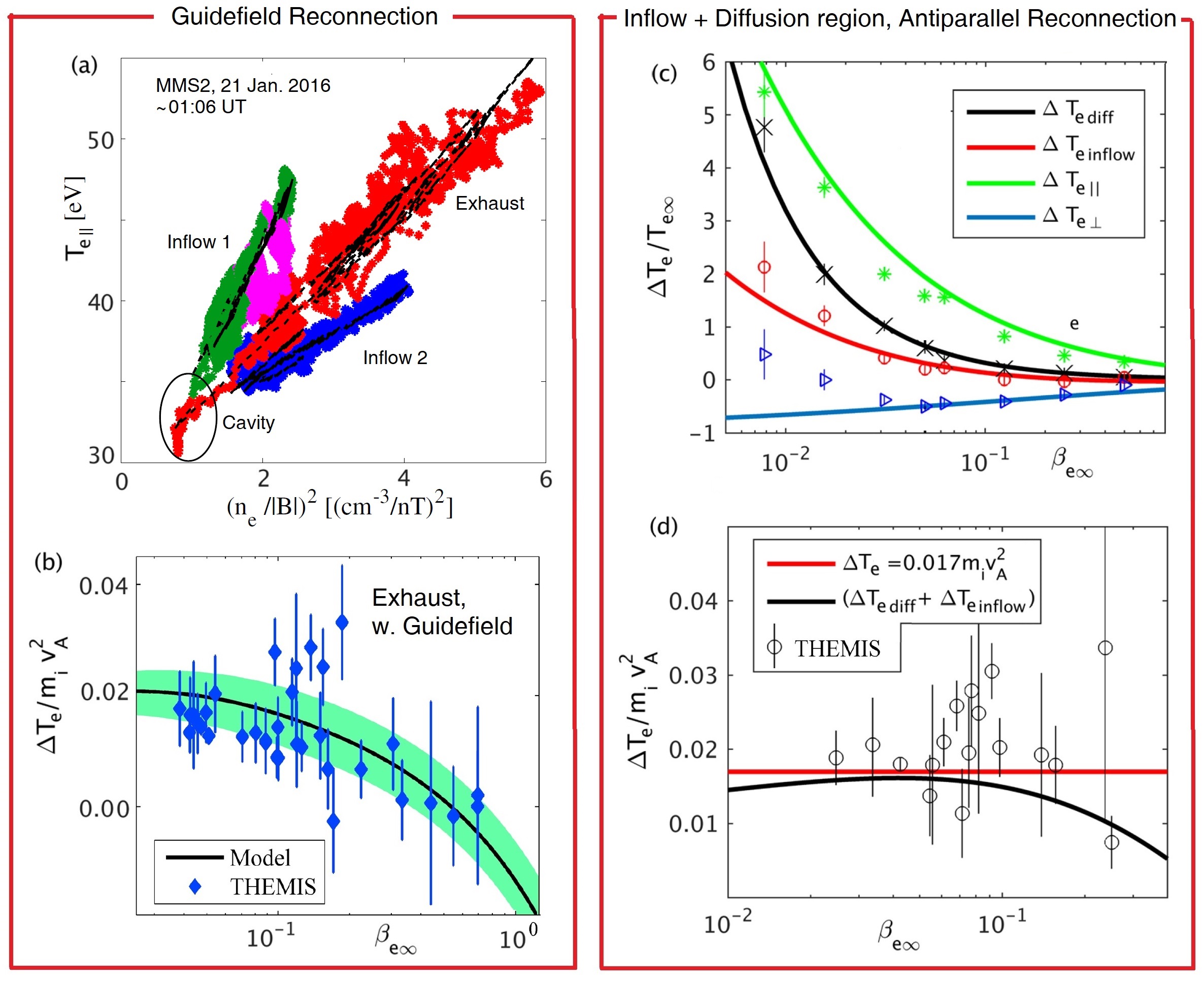}
\caption{a) MMS observations of $T_{e\|} \propto (n_e/B)^2$ for the event studied in \cite{EastwoodJP_2018, WethertonBA_2021}. b) Green band indicates the electron energization in fluid simulations \citep{OhiaO_2015} for guide-field reconnection applying the L\^{e}2009 EoS, in agreement with  observations by THEMIS \citep{PhanT_2013}. c) Analytical predictions (based on the L\^{e}2009 EoS)  for the   electron heating within the inflow and EDR of anti-parallel reconnection, validated by kinetic simulations results \citep{LeA_2016}. d) The black line show the total electron heating (sum of red and black lines in c)),  compared to results from THEMIS \citep{PhanT_2013}.}
\label{fig:Egedal_RecHeating}
\end{figure}

\mycolor{Consistent with $E_{\|} \simeq  -\nabla p_{\|}/(en)$}, \cite{EgedalJ_2013} \mycolor{found} that $e\Phi_{\|}/T_{e\infty}\propto n^2/B^2$. This dependency is much stronger than the typical Boltzmann scaling of  $e\Phi_{\|}/T_{e\infty}\propto \log(n/n_0)$ and within reconnection regions $\Phi_{\|}$ typically becomes large and is responsible for trapping and heating the majority of thermal electrons.  Fig.  \ref{fig:Egedal_Potentials} shows profiles of $\Phi_{\|}$ recorded in a range of numerical simulations. A value of $\beta_{e\infty} = nT_{e\infty}/(B_{\infty}^2/2\mu_0)\simeq 0.1$ is often applicable to reconnection within Earth's magnetosphere, yielding the profiles of  $\Phi_{\|}$ displayed in Fig. \ref{fig:Egedal_Potentials}(a, b). Meanwhile,  on occasions in the Earth's magnetotail when   lobe plasma reaches a reconnection region, the normalized pressure can drop dramatically with $\beta_{e\infty}\ll 0.1$ (see Fig.  \ref{fig:Egedal_Potentials}(d)).
 \mycolor{From the principle of quasi-neutrality, it can be shown that the required parallel streaming of electrons then exceed their thermal speed. The} dynamics then enter a non-adiabatic regime  with enhanced values of $e\Phi_{\|}/T_{e\infty} \gg 10$ over much extended spatial regions (see Fig.  \ref{fig:Egedal_Potentials}(c) as well as \cite{EgedalJ_2012, EgedalJ_2015}), likely relevant to recent MMS observations \citep{ErgunRE_2022}. In Fig.  \ref{fig:Egedal_Potentials}(e), for asymmetric reconnection the largest values of $\Phi_{\|}$ and $p_{\|}/p_{\perp}$ are observed in the low-$\beta_{e\infty}$ inflow \citep{EgedalJ_2011_popAsym, BurchJL_2016}.

During island coalescence (in Fig.  \ref{fig:Egedal_Potentials}(f) with a guide magnetic field), the effect of $\Phi_{\|}$ is also noticeable, and for this case the $p_{\|}$ values in Fig.  \ref{fig:Egedal_Potentials}(g) are enhanced by Fermi acceleration of the contracting island \citep{DrakeJ_2006, DrakeJF_2013} above the levels predicted by L\^{e}2009 EoS outlined in Fig.  \ref{fig:Egedal_Potentials}(h). \mycolor{The red line in Fig. 7(g), represents the predictions by the Le2009 EoS (see $B/B_{\infty}=1$ in Fig. 7(h)) when including Fermi heating enhancing the parallel temperature of $f_{\infty}$ by about a factor of 2.} Again, these two effects are both captured by the formalism in \cite{MontagP_2017}.

The L\^{e}2009 EoS has been verified directly by MMS during exhaust crossings of guide field reconnection, both close to \citep[$\sim 10 d_i$,][]{MontagP_2017} and far \citep[$\sim 100 d_i$,][]{WethertonBA_2021} from the $X$-line. For example, the data in Fig.  \ref{fig:Egedal_RecHeating}(a) is from the event far from the $X$-line first studied in \cite{EastwoodJP_2018}, where $T_{e\|}$ measured in the two inflows (green and blue) as well as in the reconnection exhaust (red) is \mycolor{observed to follow the  aforementioned CGL limit of the L\^{e}2009 EOS where }  $T_{e\|}  \propto  (n/B)^2$. Slightly \mycolor{a}symmetric inflow conditions set different values of proportionality,  and the exhaust comprised of a mixture of the two populations falls in the middle. The black lines represent the L\^{e}2009 EoS predictions (which also accurately account for the $T_{e\perp}$ observations, not shown here).

For guide-field reconnection, the L\^{e}2009 EoS has been implemented as a closure for the electrons in two-fluid simulations \citep{OhiaO_2012, OhiaO_2015} and as shown in Fig.  \ref{fig:Egedal_RecHeating}(b), the predicted heating levels as a function of $\beta_{e\infty}$ are consistent with THEMIS observations \citep{PhanT_2013}. Likewise, for anti-parallel reconnection, the L\^{e}2009 EoS has been applied \citep{LeA_2016} to derive theoretical scaling laws for the total electron energization as electrons approach and pass through the EDR. The theory is also consistent with kinetic simulation results as well as THEMIS observations in the reconnection exhausts (see Fig.  \ref{fig:Egedal_RecHeating}(c, d)).  \mycolor{In Fig.8(d), compared to the empirical scaling by the red line, the black theoretical curve predicts reduced heating at large $\beta_{e\infty}$. Both curves fall mostly within the error bars of the measurements. }


\begin{figure}[t]
\centering
\includegraphics[width=12cm]{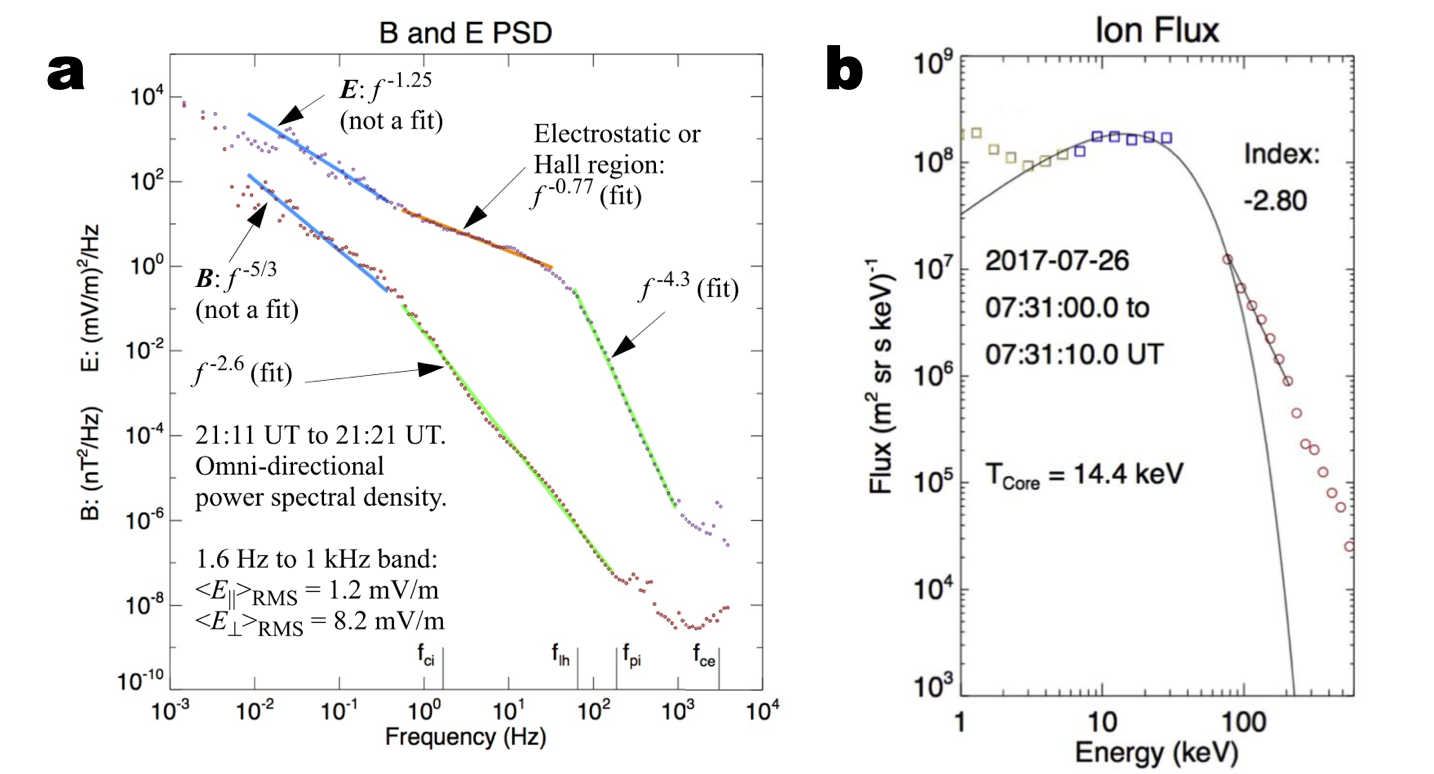}
\caption{(a) An example of $B$ and $E$ spectra in the magnetotail. The B spectrum has classic properties of turbulence, with a Kolmogorov-like inertial region (-5/3 index) and a sharp break in the region of ion dissipation. The $E$ spectra has a shallower index in the inertial region and has an electrostatic build-up at higher frequencies, before a sharp drop. The electrostatic energy density is linked to electron acceleration. Adapted from \cite{ErgunRE_2022}. (b) An example of energized and accelerated ions as measured in a region of strong turbulence. The core of the distribution is heated from $\sim$4 keV (outside of the turbulent region) to $\sim$16 keV. A high-energy tail has ions greater than 100 keV. Adapted from \cite{ErgunRE_2020_obs}.}
\label{fig:Ergun2020}
\end{figure} 

\begin{figure}[t]
\centering
\includegraphics[width=10cm]{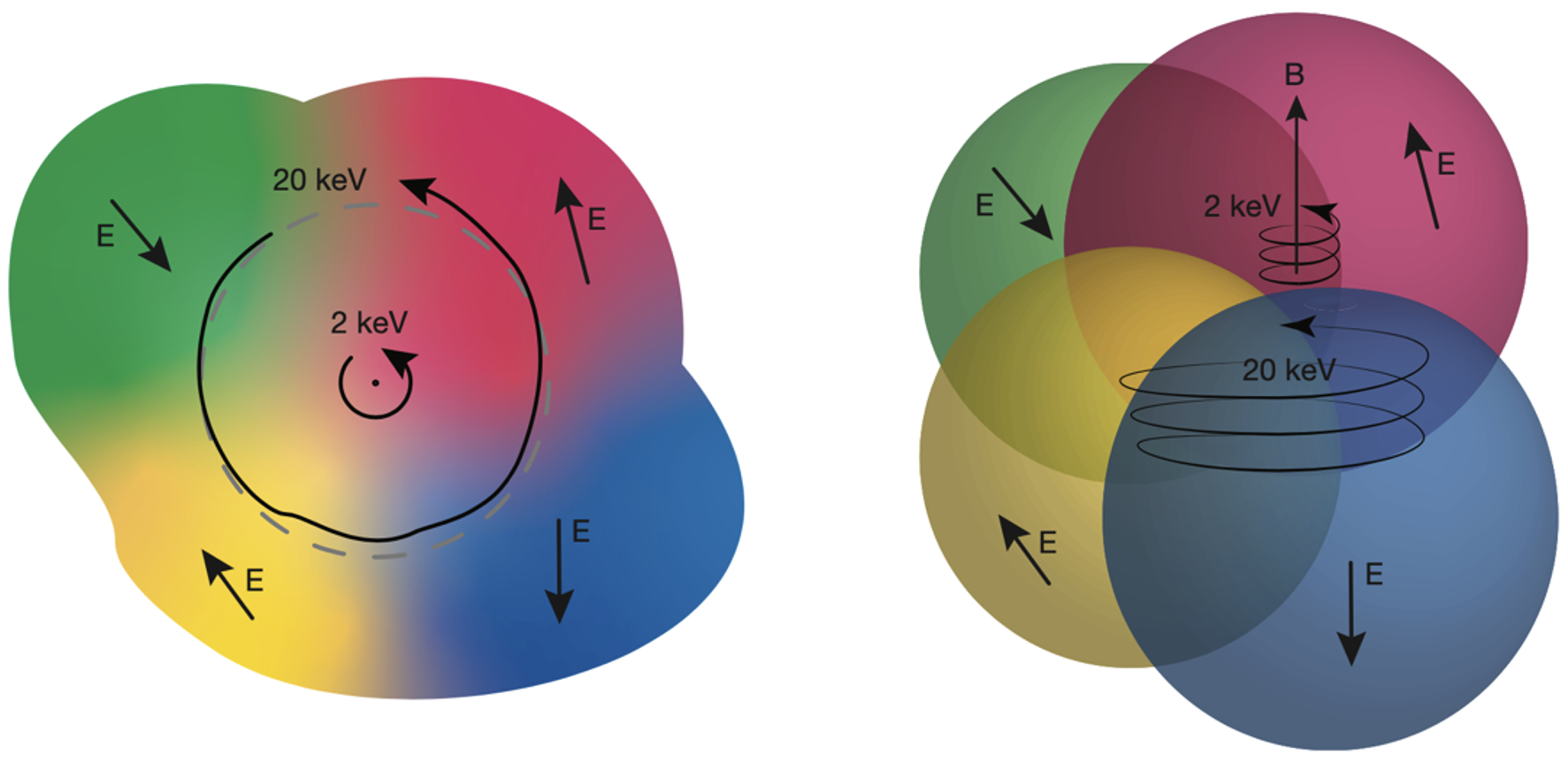}
\caption{A drawing of electron orbits in an uncorrelated, electrostatic $\mathbf{E}$, illustrating how turbulent acceleration favors higher-energy electrons. (Left) A view of the orbital plane. The higher-energy (20 keV) electron’s orbit transits several uncorrelated regions of E (including $E_{||}$) as it gyrates and, therefore, can gain or lose energy. A lower-energy electron (2 keV) sees very little change in E over an orbit. (b) A 3D view of an electron’s helical path along $\mathbf{B}$. Adapted from \cite{ErgunRE_2022_BBF}.}
\label{fig:ErgunDrawing}
\end{figure} 

\subsection{Waves and turbulence}
\label{sec:turbulence}

Magnetic reconnection in pre-existing turbulence is often referred to as `turbulent reconnection' \citep[e.g.][]{LazarianA_1999, LazarianA_2015}.  Earth's magnetosheath is such an environment, where turbulence appears to drive (smaller-scale) magnetic reconnection \citep[e.g.][]{RetinoA_2007, ErgunRE_2016_waves, PhanTD_2018}. On the other hand, magnetic reconnection can generate waves and turbulence in return \citep[e.g.][]{DaughtonW_2011, LeonardisE_2013, ErgunRE_2016_waves}. In-situ observations in Earth's magnetotail indicate that strong waves and turbulence exist in the reconnecting plasma sheet even though the upstream, lobe region is quiet, indicating that magnetic reconnection itself excites waves and turbulence \citep[e.g.][]{EastwoodJP_2009, OsmanKT_2015, ErgunRE_2018, RichardL_2023} \cite[See also][]{CattellC_1986, HoshinoM_1994}. In this section, we provide a brief review on particle energization associated with waves and turbulence observed near the X-line, including  outflow jets. Similar waves and turbulence are also found at large scales near the flow-braking region, as reviewed in Section \ref{sec:largescale}.


\subsubsection{Ion acceleration and turbulence during magnetic reconnection}

As suggested by MMS observations \citep[e.g.][]{ErgunRE_2018}, the physical process of ion and electron acceleration can differ. Because ions have larger scale sizes (skin depth and gyroradii), they are the first in line to absorb the magnetic energy. In a region of turbulence, $E$ spectra (Fig.  \ref{fig:Ergun2020}) have high enough energy density to explain the high ion energization rates though cyclotron-resonance \citep{ChangT_1986, ErgunRE_2020_sim, ErgunRE_2020_obs}. However, cyclotron resonance \mycolor{alone} does not explain an accelerated tail or other details in the ion distributions (Fig.  \ref{fig:Ergun2020}). \mycolor{Instead, a stochastic process needs to be considered, and it requires waves and turbulence that span a wide frequency range. }

\mycolor{In principle, ions can undergo Speiser-like orbits at the neutral sheet during magnetic reconnection. Nevertheless, when strong turbulence coexists within the neutral sheet, unmagnetized ions, which do not necessary follow Speiser-like orbits, are more likely to gain energy from large impulses in the turbulent electric fields. It is worth noting that}
ions with initially high energies not only absorb more powerful, larger-scale electromagnetic energy, but also have a higher probability to \mycolor{be unmagnetized and} pass through the neutral sheet. As a result, energization favors ions with initially higher energies and an accelerated tail in the ion distributions could emerge. The kinetic process of ion energization in turbulence is an active, ongoing study in which MMS observations have given good insight \citep[e.g.][]{RichardL_2022} .

\subsubsection{Electron acceleration and turbulence during magnetic reconnection}

Recent observations \citep[e.g.][]{ErgunRE_2018,LiXinmin_2022, Oka_2022} and simulations \citep[e.g.][]{LapentaG_2020,ZhangQile_2021} have provided convincing evidence that turbulence plays a significant role in accelerating electrons to non-thermal energies in the magnetotail (See Section \ref{sec:precise} for further discussion). The observations are so detailed that the specific process of interaction between the turbulence electric field and electrons can be discussed, as summarized below  \citep{ErgunRE_2018, ErgunRE_2020_sim, ErgunRE_2020_obs, ErgunRE_2022}.

Perpendicular electron energization requires circumvention of the first adiabatic invariant $(\mu=p_\perp^2/2\gamma m_0 B)$. Contrary to the case with ions, there is little power at or above the electron cyclotron frequency (Fig.  \ref{fig:Ergun2020}) and $E_{\textrm{||}}$ is small (written on plot) which suggests that electron energization should be negligible. It is found, however, that energization can occur if the correlation length scale ($d_{corr}$) in the $E$ turbulence is sufficiently small \citep{UsanovaME_2022, ErgunRE_2022}. If an electron’s parallel velocity is high enough that $d_{corr}/v_{||} < 1/f_{ce}$, it experiences changes in $E$ in less than $1/f_{ce}$ in its frame and therefore can be energized perpendicular to $\mathbf{B}$. Furthermore, if an electron's gyroradius is such that $\rho_e \geq d_{corr}$, it can experience enhanced parallel energization, perpendicular energization, and pitch-angle scattering.

Fig.  \ref{fig:ErgunDrawing} illustrates the underlying process of electron acceleration by turbulent and electrostatic $\mathbf{E}$. As it gyrates, a low-energy electron (2 keV in the figure) experiences a nearly constant E whereas a higher-energy electron (20 keV in the figure) transits regions of changing E during its gyration. Even though $\mathbf{E}$ is primarily electrostatic, the particle does not necessarily return to the same location in the perpendicular plane or in the same location along B and therefore can experience energy change. A finite $\nabla \times \mathbf{E}$ can enhance acceleration. 

The velocity dependence is such that, once again, electrons with initially higher energies are favorably energized, which results in acceleration. Interestingly, the electron energization process can be greatly enhanced by trapping in magnetic depletion \citep{ErgunRE_2020_sim, ErgunRE_2020_obs}.  Electrons can transit a turbulent region in the magnetotail in a matter of seconds, which greatly limits its energization. If trapped, the electron experiences energization for a significantly longer time, leading to much higher energization. This kinetic picture of ion an electron acceleration suggests that further study is needed.

\subsubsection{Electron energization associated with waves\label{sec:waves}}

It is instructive to discuss more specifics of what constitutes turbulence. Previous observations have shown that waves are excited over a broad range of frequency during magnetic reconnection and that they can be identified as lower hybrid waves, Langmuir waves, electrostatic solitary waves, and whistler waves \citep[e.g.][and references therein]{KhotyaintsevYV_2019}.
Perpendicular anisotropies in the region behind a dipolarization front (Section \ref{sec:largescale}) could act as a source of whistler waves \citep[e.g.][]{LeContelO_2009, KhotyaintsevYV_2011, VibergH_2014, BreuillardH_2016} or electron-cyclotron waves \citep{ZhouM_2009}.

Many studies have shown that a specific type of waves can play an important role in particle heating. For example, Debye-scale electrostatic waves and structures have been detected and discussed in the context of electron heating (or energization below $\sim$ 1 keV) near the X-line both at the magnetopause \citep{MozerF_2016, KhotyaintsevYV_2020} and the magnetotail \citep{NorgrenC_2020}. Also, an association between whistler waves and intense bursts of energetic (10s to a few 100 keV) electrons near the reconnection separatrix has been reported in the context of magnetopause reconnection \citep{JaynesA_2016, FuHS_2019a}, followed by a statistical study \citep{ChepuriSNF_2022}. \mycolor{\cite{FuHS_2019a} analyzed the energy spectrum carefully and showed that such energetic electrons are not contaminated by the magnetospheric population and yet indeed non-thermal. }


\section{Particle acceleration at large scales}\label{sec:largescale}

\subsection{Overview}
\label{sec:largescaleoverview}

While magnetic reconnection ultimately occurs at `microscopic', electron-kinetic scales within a plasma, reconnection results in macroscopic to global scale reconfiguration of the magnetic field topology and dynamics within a plasma. In the inner magnetotail, this involves inductive electric fields that are responsible for particle acceleration far removed from the actual reconnection site itself. 
The earthward exhaust region is characterized by transient or more persistent increases in the northward magnetic field, called dipolarizations. Transient events are  typically associated with rapid flow bursts, which come to rest and/or get diverted azimuthally in a `flow-braking region' near or inside of about 10 $R_E$ distance downtail. This is not a fixed distance, however. The fact that dispersionless energetic particle flux increases at tens to hundreds of keV (denoted `injections') are frequently observed at geosynchronous orbit \citep[e.g.][]{LezniakTW_1970, BakerDN_1979}, or even inside, is an indication that impulsive electric fields can often penetrate more deeply than the fast flows.

The transient dipolarization events and their associated flows are related to motional electric fields, which may exceed the electric field defining the rate of reconnection. These electric field enhancements are sometimes referred to as ‘rapid flux transport’ (RFT) events \citep[e.g.][]{SchodelR_2001_cps}. The dipolarization events typically include sharp increases of the northward magnetic field $B_z$, called `dipolarization fronts' \citep[DFs; e.g.][]{NakamuraR_2002, RunovA_2009, SitnovMI_2009}, followed by an interval of increased $B_z$, denoted `dipolarizing flux bundle' \citep[DFB;][]{LiuJ_2013} or `Flux Pileup Region' \citep[FPR;][]{KhotyaintsevYV_2011}. Further details on the terminology and properties of particle acceleration are given in recent reviews \citep{SitnovM_2019, FuHS_2020, BirnJ_2021_review}. 

Transient DFs typically separate a colder denser population in the pre-existing plasma sheet from the hotter, more tenuous population in the DFB, presumably ejected out from the reconnecting X-line \citep[e.g.][]{RunovA_2011_superposed, RunovA_2015}. Similar structures are detected for tailward flows as well (with $B_z<0$), and thus a more generalized term `reconnection front' is also used to combine both earthward and tailward cases \citep[e.g.][]{AngelopoulosV_2013}. 

Dipolarizations in the flow-braking region tend to show more persistent increases in $B_z$ \citep[e.g.][]{RunovA_2015} as well as low or decreasing earthward flow speeds, which may include tailward bounces and oscillations \citep[e.g.][]{PanovEV_2010, LiuCM_2017}. They are commonly accompanied by strong electric fields, which may exceed the motional electric field of the transient events by one or more orders of magnitude up to about 100 mV/m \citep[e.g.][]{ErgunRE_2015, ErgunRE_2022_BBF}. In contrast to the RFT electric fields, which are typically duskward, the high-frequency fields also include significant field-aligned components.

Numerous investigations have confirmed that the inductive electric fields associated with dipolarization events are the eminent cause of energetic particle flux increases, including injections observed at geosynchronous orbit. Their properties are briefly reviewed in Sections \ref{sec:anisotropies} – \ref{sec:sources}. The effects of the fluctuating strong electric fields in the flow-braking region are not as well documented. They presumably arise from the turbulence associated with the flow-braking and diversion of the earthward flow and may provide a mechanism for particle energization, separate from, or in addition to, the effects of the transient fields, and contribute a source population for the outer radiation belt \citep{ErgunRE_2022}, as well as a mechanism for energy dissipation \citep[e.g.][]{StawarzJE_2015, ErgunRE_2018}.

\subsection{Anisotropies in dipolarization events}
\label{sec:anisotropies}

\begin{figure}[t]
\centering
\includegraphics[width=12cm]{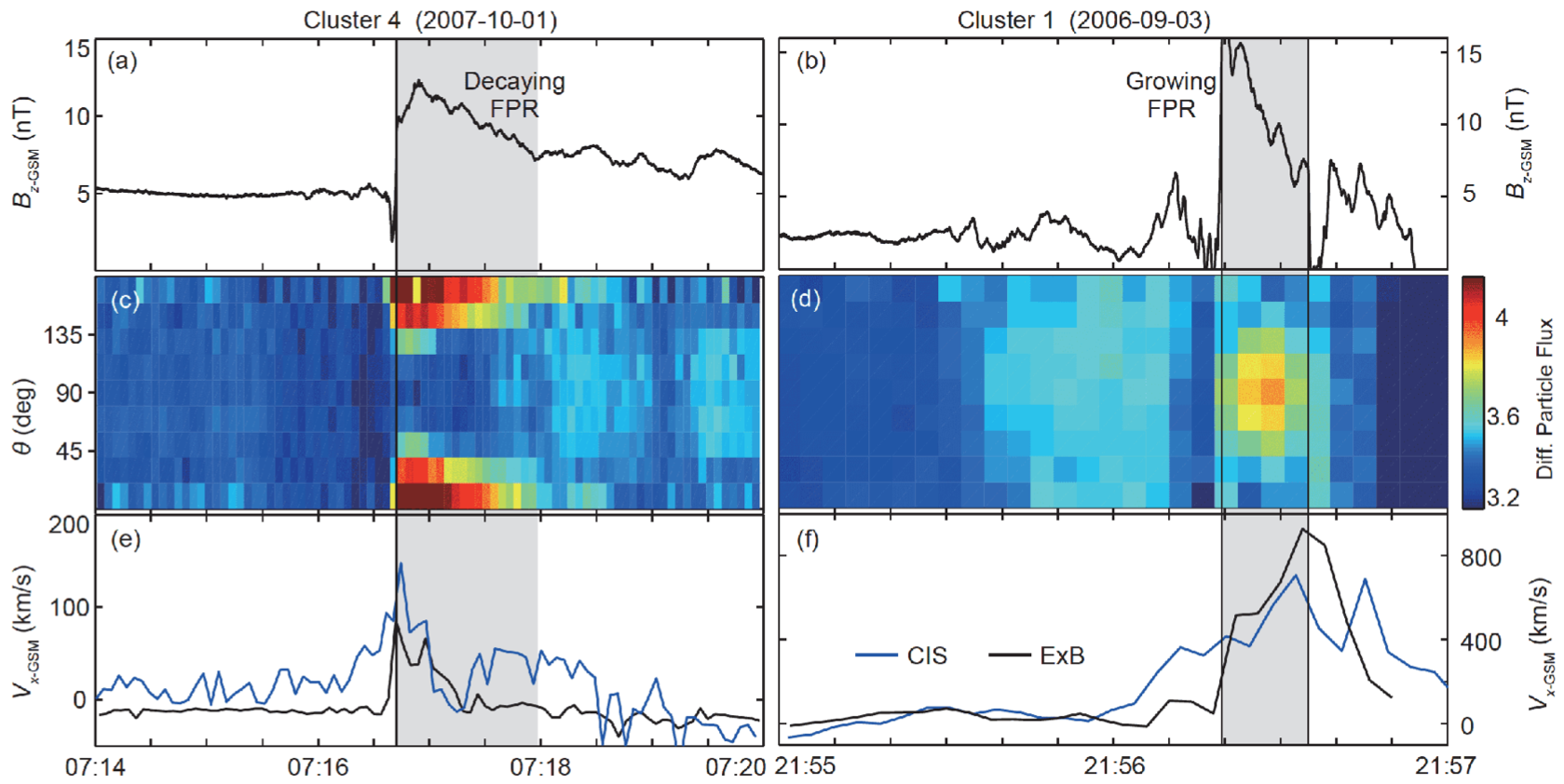}
\caption{Two different observations of dipolarization fronts, demonstrating features consistent with predominantly Fermi (left column) and betatron (right column) acceleration \citep{FuHS_2011}. When the outflow speed is decreasing (Panel (e)), the Flux Pileup Region (FPR) is considered decaying (Panel (a)) and the energetic ($>$40 keV) electrons exhibit parallel anisotropy (Panel (c)). In contrast, when the outflow speed is increasing (Panel (f)), the FPR is considered growing (Panel (b)) and the energetic electrons exhibit perpendicular anisotropy (Panel (d)).}
\label{fig:Fu2020}
\end{figure}

Many observations indicate that particles can be accelerated to non-thermal energies at and around the transient dipolarization events \citep[e.g.][]{Apatenkov_2007, RunovA_2009, FuHS_2011, FuHS_2013, Ashour-AbdallaM_2015, LiuCM_2017}, with anisotropies of the energetic particle distributions providing major clues of the underlying mechanism. Fig.  \ref{fig:Fu2020} shows two example observations by Cluster reported by \cite{FuHS_2011}.  One event was obtained when the bulk flow speed was decreasing and thus the main magnetic structure (denoted FPR, in this case) was considered decaying (left column). The other event was  obtained when the bulk flow speed was increasing, and thus the FPR was considered growing (right column). The energetic ($>$ 40 keV) electrons showed parallel anisotropy (indicating Fermi acceleration) and perpendicular anisotropy (indicating betatron acceleration) in the decaying and growing cases, respectively. 
Based on a  statistical analysis of pitch-angle anisotropy, \cite{WuMingyu_2013} consistently argued that, because outflow jets have higher speeds in the mid-tail region ($X \lesssim -15 R_E$), there could be more efficient compression of the local magnetic field, leading to more frequent formation of the perpendicular anisotropy by betatron acceleration in the mid-tail region. 

It is important to distinguish full particle acceleration, which involves the history of a particle motion, from the local acceleration rate. Estimating the latter, several investigations concluded that locally betatron acceleration was dominant at the \mycolor{dipolarization front (DF)} proper \citep[e.g.][]{XuY_2018, FuHS_2019b, MaWenqing_2020} and that the Fermi acceleration would be more effective at a larger spatial scale. Using MMS data, \cite{MaWenqing_2020} showed that betatron acceleration rate dominates at many dipolarization fronts in the magnetotail in the $X < -10$ R$_E$ range. Such a conclusion is consistent with the earlier, global-scale picture in which electrons are expected to experience predominantly Fermi acceleration in the stretched magnetic field in the magnetotail but undergo betatron acceleration as the magnetic field increases \citep[e.g.][]{SmetsR_1999}.

\begin{figure}[t]
\centering
\includegraphics[width=12cm]{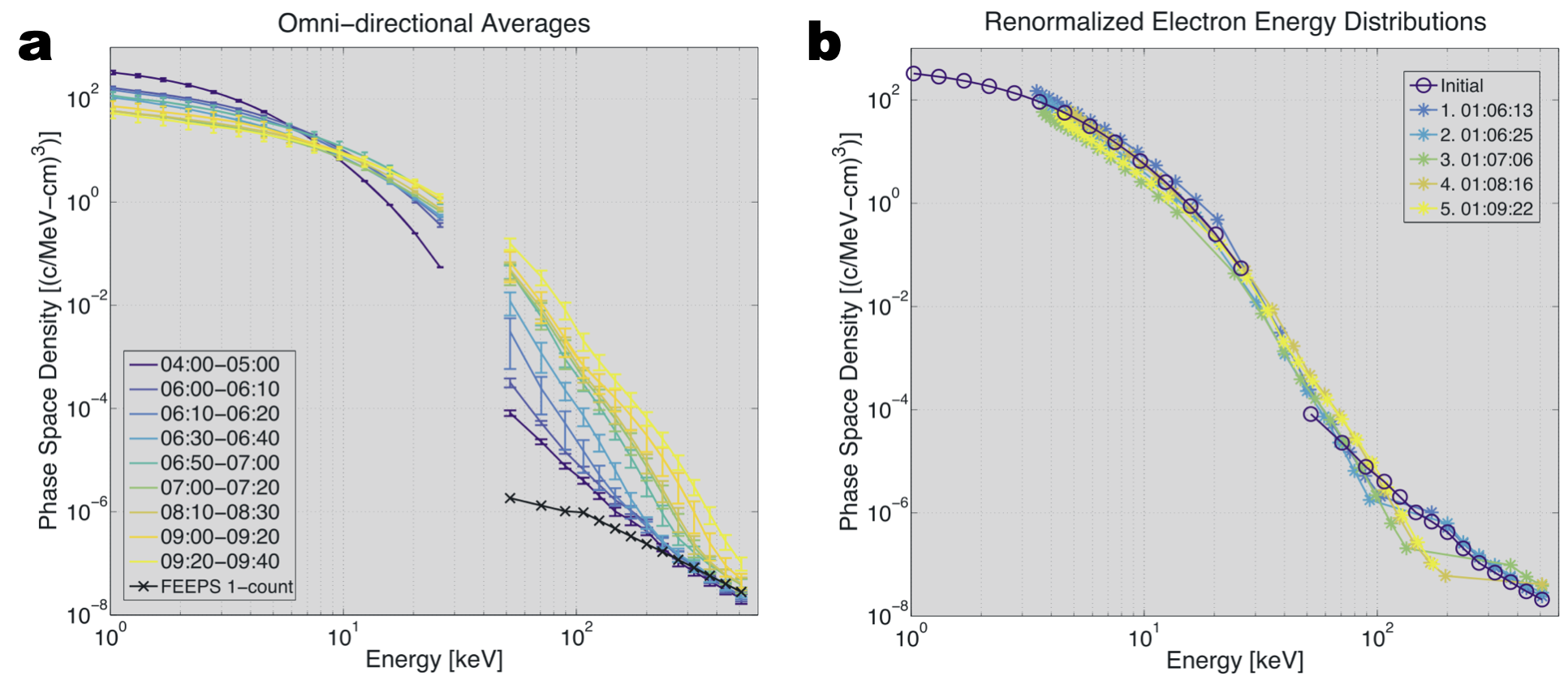}
\caption{Electron energy spectra obtained by MMS, demonstrating the importance of the betatron process. (a) Electron energy spectra observed during a series of dipolarization events. The times are listed in minutes and seconds (mm:ss); all of them have the same hour, 01:00 UT. (b) The same energy spectra but renormalized using the simple model that assumes betatron acceleration. See \cite{TurnerDL_2016} for more details. }
\label{fig:Turner2016}
\end{figure}

\cite{TurnerDL_2016} also used observations from NASA’s MMS mission to demonstrate how electron acceleration associated with a dipolarization structures and BBFs in the magnetotail were energy-dependent but consistent with betatron acceleration (Fig. \ref{fig:Turner2016}). \cite{MalykhinAY_2018} examined 13 dipolarization events using Cluster data and concluded that the electron acceleration up to 90 keV was consistent with betatron acceleration. \cite{VaivadsA_2021} used Cluster data in the magnetic structures (flux rope and dipolarization) of an Earthward reconnection jet, and found that in the dipolarization structure, electron acceleration was generally consistent with betatron acceleration, while within the flux rope, electron acceleration was more consistent with Fermi acceleration. While most conclusions are from single point measurements, \cite{NakamuraR_2021} used a constellation between MMS and Cluster satellites to infer consistency with adiabatic acceleration of electrons trapped within a dipolarization structure.

Electron anisotropies may vary not only with distance from the Earth or from the reconnection site but also with respect to the distance from the neutral sheet ($B_x \sim 0$). \cite{RunovA_2013} reported pancake type distributions (90$^{\rm o}$ peaked) near the neutral sheet and mostly cigar type (0$^{\rm o}$ and 180$^{\rm o}$ peaked) distributions away from the neutral sheet, consistent with a predominance of betatron acceleration of $\sim$90$^{\rm o}$ particles close to the neutral sheet and Fermi acceleration for field-aligned electrons reaching higher latitudes.

Ions can also be accelerated in association with BBFs and dipolarization events, as studied by recent MMS observations \citep[e.g.][]{BinghamST_2020, RichardL_2022} and simulations \citep[e.g.][]{ParkhomenkoE_2019, BirnJ_2015_DF}. Because the ion gyro-radii are relatively large, they do not conserve the adiabatic moment, except in some average sense, and often behave non-adiabatically.  Ion acceleration is further discussed in Section \ref{sec:mechanisms}.

\subsection{Acceleration mechanisms}
\label{sec:mechanisms}

\subsubsection{Electrons\label{sec:electrons}}

The spatially and temporally localized cross-tail electric field associated with earthward propagating dipolarization fronts can result in trapping, earthward transport, and rapid acceleration of energetic particles, leading to the betatron effect from drift toward increasing $B$-fields. Various models, which capture the essential localization of the $E$-field, have been based on the adiabatic drift approximation, concentrating on equatorial drift orbits. They clearly demonstrated how the motional, azimuthally oriented electric field associated with a magnetotail dipolarization and corresponding bursty bulk flow (BBF) of 100s km/s can accelerate energetic particles and transport them rapidly radially inward with the BBF itself \citep[e.g.][]{LiXinlin_1998, GabrielseC_2012, GabrielseC_2014, GabrielseC_2016, GabrielseC_2017}, and yield flux increases consistent with energetic particle observations.


The localized electric field in RFT events can also cause parallel Fermi acceleration of ions and electrons bouncing through this region once or (for electrons) multiple times. Studies of this effect require orbit tracing in three-dimensional magnetic and electric fields, which are usually obtained from MHD simulations \citep[e.g.][]{BirnJ_1994, BirnJ_2004, Ashour-AbdallaM_2011, SorathiaKA_2017}. These studies confirmed the mechanism of temporal magnetic trapping within the magnetic field structures of DFBs, not only for electrons but also for ions \citep{BirnJ_2015_DF, UkhorskiyAY_2017, UkhorskiyAY_2018}, and  showed the rapid acceleration via betatron (perpendicular to the B-field) and/or Fermi (parallel to B) effects (Section \ref{sec:GCA}). They demonstrated not only parallel and perpendicular anisotropies of energetic electron distributions, but also so-called `rolling pin' distributions \citep{LiuCM_2017_rollingpin} with peaks at 0$^{\rm o}$, 90$^{\rm o}$, and 180$^{\rm o}$ pitch angles \citep{RunovA_2013, BirnJ_2014, BirnJ_2022_anisotropy}, depending on energy, time and location.


Here, it is worth emphasizing again that even electron motion is not necessarily always adiabatic, especially at and around the X-line, in strongly curved low-$B$ fields, or in regions of strong waves and turbulence, e.g. near the reconnection site. In such cases, the parallel electric field carried by whistler waves  (Section \ref{sec:turbulence}) or  kinetic Alfv\'{e}n waves \citep[e.g.][]{GuoZhifang_2017} might be important in addition to Fermi and betatron acceleration.

The fate of DFBs and associated energetic particles has also been investigated within the Rice-Convection-Model \citep[RCM-E; ][]{ToffolettoF_2003}, covering the energy-dependent drift of depleted magnetic flux tubes (also denoted `bubbles') within a quasi-static inner magnetosphere model \citep{YangJian_2013, YangJian_2015}. In this regard, accelerated electrons were demonstrated to be important as a likely seed population of the Earth’s radiation belt.  \cite{SorathiaK_2018} conducted test-particle simulations of electrons in high-resolution, dynamic MHD fields to show how energetic electron injections from the magnetotail likely contribute a significant and possibly even dominant source of outer radiation belt electrons in the 100s of keV range in the inner magnetosphere. \cite{TurnerDL_2021_source} conducted a phase space density analysis using a combination of Van Allen Probes in the outer radiation belt and MMS in the magnetotail plasma sheet to demonstrate also that relativistic electron acceleration in the plasma sheet can result in sufficient intensities to serve as a direct source for outer radiation belt electrons. \mycolor{Here, it is worth emphasizing that the intensities in the magnetotail can get up to radiation belt levels yet the residence time of those electrons in the tail is only a few minutes, in contrast to the several days residence times in the outer radiation belt.}



\subsubsection{Ions\label{sec:ions}}

Acceleration of ions in dipolarization events can be similar to that of electrons. Details are summarized in recent reviews by \cite{SitnovM_2019} and \cite{BirnJ_2021_review} with references therein. Simulations by \cite{BirnJ_2015_beam} showed how the acceleration of protons in the \mycolor{central plasma sheet (CPS)} is generally consistent with the betatron effect (with an average conservation of the first adiabatic invariant in the presence of an increase in magnetic field strength).  This is consistent with conclusions of \cite{UkhorskiyAY_2017, UkhorskiyAY_2018}, which were based on test particle tracing in high-resolution global MHD simulations. The simulations also demonstrated parallel acceleration (similar to Fermi acceleration of type B) by single (or, in rarer cases, multiple) encounters of a dipolarization front. In contrast to electrons, a single encounter of, or reflection at, a dipolarization front may result in observable, albeit moderate-energy proton beams or precursor populations preceding a DF \citep[e.g.][]{ZhouXZ_2010, ZhouXZ_2011, BirnJ_2015_DF}.  Presumably, a similar process can also happen at reconnection fronts on the tailward/anti-earthward side of magnetotail reconnection. Due to the mass dependence \mycolor{of the gyroradius that characterizes the encounter or reflection}, this energization is even more effective for heavier ions, such as oxygen. As the energy gain essentially results from picking up the speed of the moving structure, it was also likened to a `pick-up' process \citep{DelcourtDC_1994, EastwoodJP_2015, BinghamST_2021, BirnJ_2021_Oxygen1_CPS}.


The simulations have yielded characteristics of ion distributions, dominated by protons, that are consistent with observations right after passage of a DF. At low distance from the neutral sheet, in the central plasma sheet, distributions show perpendicular anisotropy \citep{RunovA_2015, RunovA_2017, BirnJ_2017_CPS, ZhouXZ_2018}, consistent with the betatron effect, which may be accompanied by lower intensity, lower energy field-aligned counter-streaming beams. At larger distance from the neutral sheet, close to the plasma sheet boundary, the distributions consist of crescent-shaped earthward field-aligned beams \citep[e.g.][]{ZhouXZ_2012_emergence}. \mycolor{At the distance slightly away from the plasma sheet boundary layer (PSBL) and closer to the neutral sheet, such crescent-shaped earthward beam can be }accompanied by tailward beams, which apparently result from mirroring closer to Earth. \mycolor{In such a region}, sometimes multiple earthward and tailward beams are observed, which may be considered the counterparts of the field-aligned electron populations, however, involving only few bounces \citep{BirnJ_2017_PSBL}. It is noteworthy that crescent-shaped earthward ion beams (including their tailward streaming counterparts) can also result from reconnection deeper in the tail \citep{AndrewsMK_1981_jetting, ForbesTG_1981_retreat, WilliamsDJ_1981_ionbeams}.

At higher energies, or for heavier ions, the gyroradius becomes comparable to, or larger than the size of the dipolarizing acceleration region, and the ions may encounter this region exhibiting Speiser-type orbits or even traverse the acceleration region of the enhanced electric field in a demagnetized fashion \citep{BirnJ_2021_Oxygen1_CPS, RichardL_2022}. \mycolor{The energy gain is essentially given by
\begin{equation}
    \Delta W = q \int E_y dy
\end{equation}
where $q$ and $E_y$ is the particle charge and the enhanced electric field \citep{BirnJ_2021_Oxygen1_CPS}.} This provides an upper limit to the possible acceleration of a given species, which is higher for multiply-charged ions. In agreement with that conclusion, the $E/q$ dependence of particle fluxes in MMS observations of energetic particle events associated with fast flows \citep{BinghamST_2020, BinghamST_2021} indicated that He$^{++}$ and O$^{6+}$ of solar wind origin dominated the particle fluxes at highest energies ($>$ 400 keV). The non-adiabatic acceleration effects also lead to non-gyrotropic, phase bunched, velocity distributions of heavy ions \citep{DelcourtDC_1997, BirnJ_2021_Oxygen1_CPS}.


\subsection{Sources and seeding}
\label{sec:sources}

\begin{figure}[t]
\centering
\includegraphics[width=12cm]{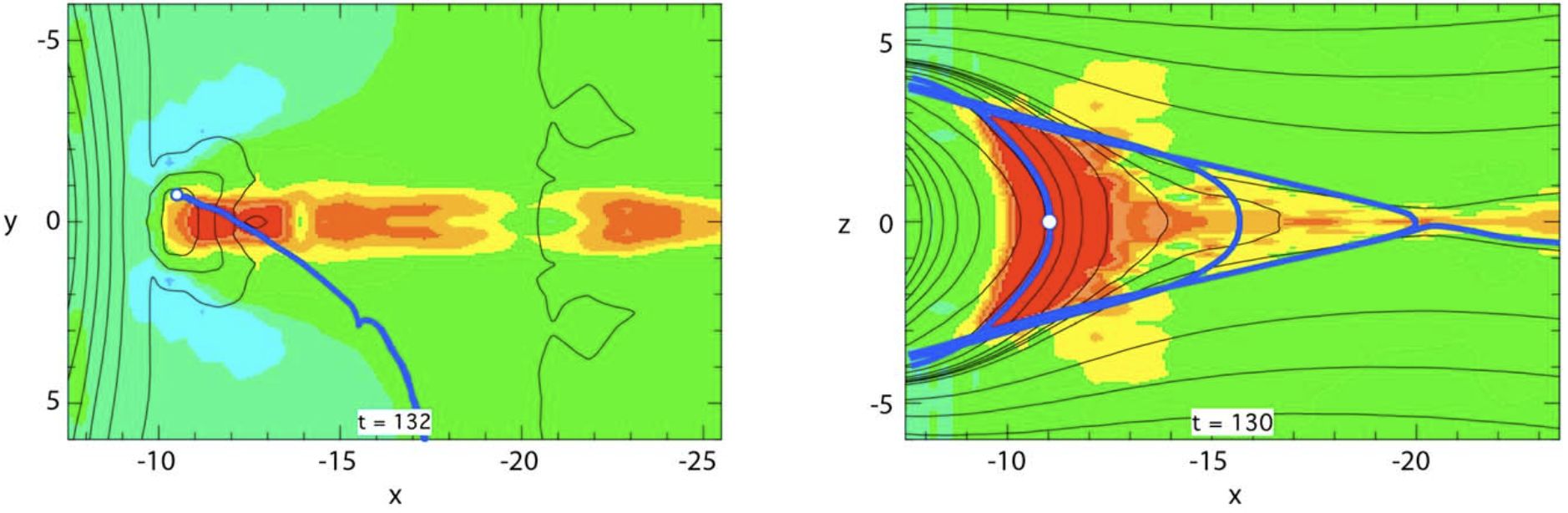}
\caption{Simulation results demonstrating two main sources of accelerated electrons: (a) near-equatorial drift from the dusk flank which leads to the energization by the localized motional electric fields and (b) a bounce orbit that originates from the reconnection region and leads to Fermi and betatron acceleration. Adapted from \cite{BirnJ_2014}.}
\label{fig:Birn2014}
\end{figure}

Observations do not give direct information about the sources of the accelerated particles. The fact that transient DFs typically separate a hotter, more tenuous population inside a DFB from the colder, denser plasma ahead of it indicates that the pre-DF population is not the source of the energized population inside the DFB. More definite conclusions about the sources come from modeling, particularly from particle tracing in fields modeling the inward propagation of DFBs.   Through backward tracing in dynamic MHD fields,  \cite{BirnJ_2012, BirnJ_2014} demonstrated how particles are seeded onto the reconnected field lines inside a DFB, thus gaining access to the acceleration processes, via two mechanisms: i) local cross-tail particle drifts in the plasma sheet configuration, and ii) direct entry enabled by remote reconnection of field lines (Fig. \ref{fig:Birn2014}). The entry mechanisms are energy-dependent: at low energies, charged particles are closely tied to the field lines that undergo reconnection before participating in the inner tail collapse, whereas at higher energies, cross-tail drifts or even non-adiabatic cross-tail motions become more important and particles can enter the acceleration region from the flanks earthward of the reconnection site. 

\cite{TurnerDL_2016} examined MMS observations of a series of dipolarizations associated with magnetotail reconnection and found that for electrons with energy $>$10 keV, extending into the relativistic range, the observed acceleration was largely consistent with betatron acceleration, and one important consequence of those observational results was that the source of electrons in the ambient, background plasma sheet must have been relatively uniform over a large portion of the magnetotail surrounding the MMS spacecraft. The upper energy limitation of electron flux increases found by \cite{TurnerDL_2016} and earlier by \cite{BirnJ_1997b} is consistent with the change of particle motion and source regions at high energies mentioned above.

Using again MHD/test particle simulations, \cite{BirnJ_2022_anisotropy} further demonstrated energy and space dependence of source regions of accelerated electrons. Consistent with earlier conclusions, they explained the drop in fluxes observed at energies $<$ 10 keV (consistent also with the results of \cite{TurnerDL_2016}) as being related to the drop in density from the seed populations in the \mycolor{plasma sheet boundary layer (PSBL)} and lobes, despite the fact that these particles were also adiabatically accelerated.

\begin{figure}[t]
\centering
\includegraphics[width=12cm]{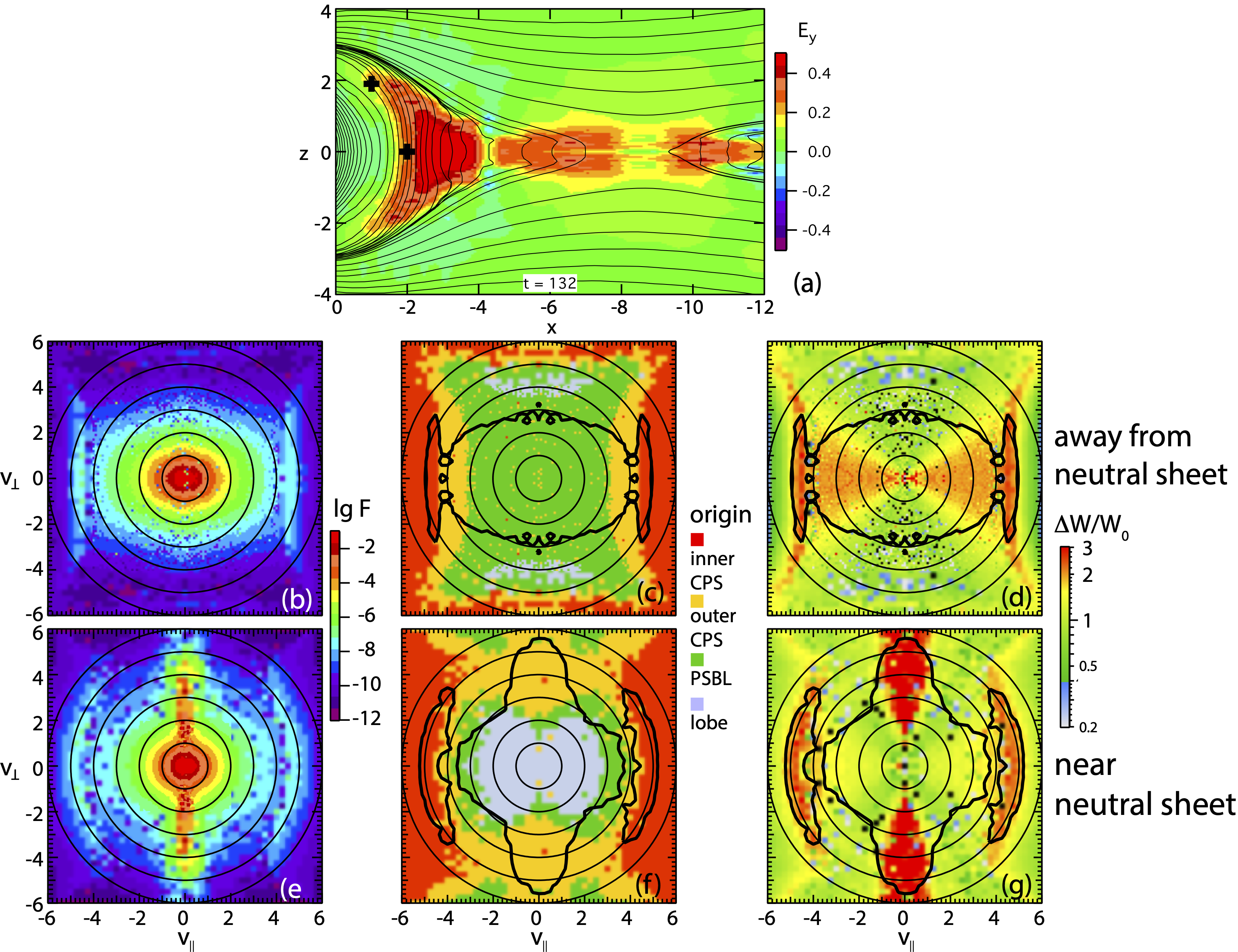}
\caption{Electron pitch angle distributions (PADs) obtained from a modeling approach \citep[modified after ][]{BirnJ_2022_anisotropy}. The top panel shows a snapshot of the cross-tail electric field of the underlying MHD simulation, obtained right after the arrival of a dipolarization front; crosses indicate two locations where PADs were obtained (panels b-g). Panels b and e show the PADs at the two locations. Panels c and f indicate the origins of the particles contributing to the PADs, and panels d and g show the relative energy gain along each particle trajectory. Based on the chosen parameters, the velocity unit corresponds to $(m_p/m_e)^{1/2}$ 1000 km/s = 42,850 km/s, or an energy of 5.2 keV.}
\label{fig:Birn2022}
\end{figure}

Figure \ref{fig:Birn2022}, modified after Fig. 4 of \cite{BirnJ_2022_anisotropy}, illustrates some important conclusions from modeling electron pitch angle distributions (PADs) right after the passage of a DF. The MHD configuration is indicated in the top panel (a). 
\begin{enumerate}
    \item Panels b and e demonstrate characteristic anisotropies of cigar-type (field-aligned) and pancake-type (perpendicular) away from, and close to, the neutral sheet, respectively; the two locations are indicated by the crosses in panel a. This result agrees with observations by Runov et al. (2013). 
    \item Panels c and f show the origins of the particles contributing to these PADs, demonstrating that they are composed of different sources: At the highest energies particles originate from the inner CPS, as illustrated in Fig. 13a, whereas at lower energies the outer CPS, the PSBL, and the lobes contribute, as shown in Fig. \ref{fig:Birn2014}b. 
    \item Panels d and g show the relative energy gain along the phase space trajectory, represented by the ratio between the final energy and the energy at the source location. These panels illustrate the effects of `heating', increasing particle energies by a similar factor over a wide energy range, versus the acceleration of particles in a limited energy (and pitch angle) range. The distribution away from the neutral sheet in panel d shows the Fermi ‘heating’ at pitch angles around 0 and 180 degrees, whereas the distribution near the neutral sheet in panel g shows the betatron heating near 90$^{\rm o}$ pitch angles. Both panels show an accelerated field-aligned population at $v\sim$4 - 5 (corresponding to $\sim$80 - 130 keV for the chosen units), although in panels b, d this is distinct from the `heated' population mainly by the source locations in the inner CPS, where densities are higher. 
    \item \mycolor{The three peaks near 0$^{\rm o}$, 90$^{\rm o}$, and 180$^{\rm o}$ in panels e and g}  also illustrates the formation of the `rolling-pin' distribution, documented observationally \citep[e.g.][]{RunovA_2013, LiuCM_2017_rollingpin}. \mycolor{It is a combination of dominantly parallel (`cigar'-shaped) and perpendicular (`pancake'-shaped) distributions.}
\end{enumerate}

The model particle tracing provides information on the immediate source regions, such as plasma sheet vs. lobes, which cannot easily be inferred from observations. Ultimately, particles originate from two sources, the solar wind and the ionosphere. The distinction between the two source regions was made traditionally on the basis of ion composition experiments, with H$^+$ indicating solar wind origin, while the presence of O$^+$ indicated an ionospheric source \citep[e.g.][]{ShelleyEG_1972}. This view has been extended and modified significantly. On one hand, detailed test particle tracing studies in global MHD models of storm time magnetosphere evolution have demonstrated that ionospheric H+ can also populate the plasma sheet and provide a seed population \citep{GlocerA_2020}. On the other hand, detailed studies of the energy/charge dependence of enhanced energetic particle fluxes showed that the contribution of heavy energetic ions to enhanced fluxes is not a fixed percentage but rather depends on energy and charge status, with O$^+$ (of ionospheric origin) dominating at lower energies of tens of keV, while multiply charged oxygen, particularly O$^{6+}$ of solar wind origin, was found to dominate at energies of hundreds of keV \citep{CohenIJ_2017, BinghamST_2020, BinghamST_2021}. 

Reconnection on the dayside might also contribute to seeding of energetic particles in the near-Earth space environment. \cite{FennellJF_2016} reported on `microinjections' of relativistic electrons observed by MMS; microinjections are frequent and rapid, energy-dispersed to dispersionless enhancements of electron intensities observed around $\sim 10 R_E$ geocentric distance along the tailward-flanks of the magnetosphere. By tracing dispersed particle signatures back to their dispersionless origins, \cite{FennellJF_2016} demonstrated that dispersed microinjection observations along the dusk-side of the magnetosphere map back to near the subsolar and early afternoon magnetopause. \cite{KavosiS_2018} showed that the observed periodicity of microinjection electrons is consistent with a combination of Kelvin-Helmholtz (KH) waves and \mycolor{flux transfer events (FTEs)} along the dayside magnetopause. Those results indicate that microinjected electrons might result from bursts of reconnection associated with KH instability and FTEs along the dayside magnetopause. Conversely, the drops in fluxes around microinjection electrons might also be the signature of losses of energetic electrons through the magnetopause and to the magnetosheath; however that electron loss process too is only enabled via reconnection resulting in magnetic connectivity across the magnetopause \citep[e.g.][]{KimKC_2014, MaukB_2016_escape}. Tracing test particles in a dynamically evolving MHD model has reproduced even the salient features of losses \mycolor{(including detailed variations both in space and time and the depth of penetration and persistence of particles in the magnetosheath)} for different species in agreement with MMS observations \citep[e.g.][]{SorathiaKA_2017}.

\subsection{Diamagnetic cavities}
\label{sec:diamagnetic}

An interesting topic that drew some attention in the recent decade is the diamagnetic cavities that form at high magnetic latitudes in the cusp region as a consequence of large-scale, dayside magnetopause reconnection \citep[e.g.][]{LavraudB_2002, LavraudB_2005}.
This region has a substantially reduced magnetic field magnitude and is filled with dense, sheath-like plasma with high-energy ($>$ 30keV) electrons and ions (including heavy ions). The high-energy particles exhibit perpendicular anisotropy \citep[e.g.][]{NykyriK_2019}, and test-particle simulations suggest that those high-energy particles are produced locally via betatron and/or Fermi mechanisms while being trapped in the magnetic bottle like configuration associated with the cavity \citep{NykyriK_2012, NykyriK_2019, SorathiaKA_2019, BurkholderBL_2021}. The relatively large size of the diamagnetic cavities, i.e., 3-5 $R_E$ in width \citep{NykyriK_2019} indicates that they can be a major source of plasma (electrons, protons and oxygen ions) into Earth’s magnetosphere as well as providing a high-energy particle source \citep{NykyriK_2021}. 

\section{Outstanding Problems}
\label{sec:discussion}

There remain unsolved problems in the topic of particle acceleration by magnetic reconnection in geospace. Here, we describe two topics, energy partition and the precise role of turbulence. These problems are very relevant to particle acceleration in solar flares. 

\subsection{Energy partition}
\label{sec:partition}

For solar flares, it has been reported that non-thermal electrons alone carry up to 50\% of the released magnetic energy \citep[e.g.][]{LinRP_1976, AschwandenMJ_2017}. In fact, more detailed studies argue that thermal electrons can indeed carry much less energy than non-thermal electrons, even in coronal sources \citep[e.g.][]{KruckerS_2010, KruckerS_2014, FleishmanGD_2022}. This is in stark contrast to the case of Earth's magnetotail (in particular the reconnection region) where non-thermal electrons appear to carry only a minuscule fraction of released energy \citep[e.g.][]{OierosetM_2002}. 

While the plasma parameters in the magnetotail differ greatly from those in the solar atmosphere, it is still instructive to know how energy is partitioned between thermal and non-thermal components in the magnetotail.  A caveat is that the typical particle energy spectrum in the magnetotail does not exhibit a clear spectral break, and it is difficult to separate those components at a certain energy $E_c$ \citep[e.g.][]{ChristonSP_1988, ChristonSP_1989, ChristonSP_1991,OierosetM_2002, OkaM_2018}. Fortunately, the energy spectrum is often well approximated by the kappa distribution and the non-thermal fraction of particle energy (and also density) can be calculated analytically without introducing a sharp boundary at $E_c$ \citep{OkaM_2013, OkaM_2015}. 

Based on the kappa distribution model, it was shown that, for the above-the-looptop (ALT) hard X-ray coronal sources in solar flares, the fraction of non-thermal electron energies was at most $\sim$50\%, indicating equipartition between thermal and non-thermal components \citep{OkaM_2013, OkaM_2015}. Similar values of non-thermal fraction were obtained by self-consistent particle simulations of magnetic reconnection \citep[e.g.][]{ArnoldH_2021, ZhangQile_2021}, as well as {\it in situ} observations of electron energy spectra during magnetotail reconnection \citep{Oka_2022}. 

A puzzle is that, even when electrons are  significantly heated (for example, the event of 2017 July 26, Fig. \ref{fig:example_events} right), the non-thermal tail does not necessarily become harder \citep{RunovA_2015, ZhouMeng_2016, Oka_2022}. This is counter-intuitive because the non-thermal tail is often expected to be enhanced as the temperature increases. When electrons are not significantly heated (for example, the event of 2017 July 11, Fig. \ref{fig:example_events} left), the non-thermal tail becomes harder (softer) as the spacecraft approaches toward (moves away from) the X-line. The two distinct types of reconnection events, i.e., less heated and much heated events, can be interpreted by the concept of `plasma sheet reconnection' and `lobe plasma reconnection', respectively \citep[e.g.][and references therein]{Oka_2022}. However, it remains unclear, at least from the observational point of view, what controls the energy partition between thermal and non-thermal components of electrons. One caveat that has to be considered in the magnetotail events is that the particle distribution observed prior to an event is generally not (or not identical to) the source of the population observed afterward, as discussed in Section \ref{sec:sources}.  

For ions, the energy partition between thermal and non-thermal components is much less studied in the magnetotail, although ions do form a clear power-law tail in the magnetotail \citep[e.g.][]{ChristonSP_1988, ChristonSP_1989, ChristonSP_1991, OierosetM_2002, ErgunRE_2020_obs}. Recent particle simulations of magnetic reconnection have shown that ions and electrons form a very similar power-law tail, but non-thermal protons gain $\sim 2 \times$ more energy than non-thermal electrons \citep{ZhangQile_2021}. It was argued that the primary mechanism of acceleration is Fermi acceleration and that the strong field-line chaos associated with the flux-rope kink instability allows particles to be transported out of flux ropes for further acceleration. It is to be noted that energetic ions in the magnetotail especially in the dipolarization region can have multiple sources and thus the process of energy partition might be a little more complex. \cite{BirnJ_2015_DF} argued that an enhanced flux of energetic ions can result from not only acceleration of thermal ions in the reconnection region but also a drift entry of pre-energized ions from the magnetotail flanks. Observational validation of these scenarios of ion acceleration and associated partition of energy is left for future work.

\subsection{The precise role of turbulence}
\label{sec:precise}

Many theoretical and simulation studies have shown that the guiding-center approximation   is effective in explaining particle acceleration during magnetic reconnection and that particle acceleration is achieved by a Fermi-type mechanism involving curvature drift (Section \ref{sec:theories}). However, turbulence may also play a significant role (Section \ref{sec:turbulence}), although its importance and specific role in particle acceleration are not fully understood, at least from an observational standpoint. For example, \cite{ErgunRE_2020_sim} argued theoretically that the turbulence with high-frequency electric fields in a magnetic depletion region can energize electrons up to non-thermal energy. \cite{ZhangQile_2021} have demonstrated that the flux-rope kink instability leads to strong field-line chaos, allowing particles to be transported out of flux ropes for further acceleration by other flux ropes. Also, \cite{FujimotoK_2021} have shown that the turbulence-induced electric field at the core of flux ropes can scatter electrons, resulting in heating rather than acceleration to non-thermal energy. The turbulence in these theoretical models has different roles, and such roles have not been fully explored in observational studies. 

It is to be noted again that an enhanced turbulence may not necessarily lead to an enhanced non-thermal tail in the reconnection region (Sections \ref{sec:examples} and \ref{sec:partition}), although turbulence appears correlated with enhancements of non-thermal tail in the flow braking region \citep[e.g.][and references therein]{ErgunRE_2022_BBF}.  Also, hard electron spectra have been found even in a quiet-time plasma sheet (Section \ref{sec:active}), raising a question whether turbulence can confine electrons. After all, what makes reconnection more turbulent 
and how important turbulence is for the process of particle acceleration remain unsolved.

\section{Summary and Conclusion\label{sec:summary}}

In the past decade, a key theme of particle acceleration studies was whether the guiding-center approximation can describe particle acceleration and which of the key mechanisms, i.e., Fermi acceleration, betatron acceleration, and the direct acceleration by parallel electric field, is more dominant. The MMS mission has enabled the evaluation of each term and supported the earlier idea that both Fermi and betatron acceleration are important in many cases of electron acceleration during reconnection. In the collapsing region where the intrinsic dipole field becomes more important, the betatron acceleration dominates in the central plasma sheet. While some populations originate from the flank of the magnetotail without much increase in energy, other populations experience energization at localized dipolarization while being transported earthward from the reconnection region. In addition to the Fermi and betatron acceleration, a parallel potential develops near the reconnection X-line and traps incoming electrons, resulting in a significant \mycolor{energization}. The electric field associated with turbulence can also accelerate electrons but such process might invalidate the assumption of adiabatic particle motion. Ions are more likely to behave non-adiabatically even near Earth, away from the reconnection region. Outstanding problems remain regarding, for example, energy partition between thermal and non-thermal components and the precise role of turbulence in the particle acceleration process. Solving these problems might be helpful for understanding the particle acceleration mechanism in other plasma environments, such as the solar corona.


\backmatter
\bmhead{Acknowledgments}
We thank Seiji Zenitani for proofreading the manuscript prior to submission. This work was initiated and partly carried out with support from the International Space Science Institute (ISSI) in the framework of a workshop entitled `Magnetic Reconnection: Explosive Energy Conversion in Space Plasmas', led by Rumi Nakamura and James L. Burch.

\section*{Declarations}

\bmhead{Funding}
MO was supported by NASA grants 80NSSC18K1002, 80NSSC18K1373, and 80NSSC22K0520 at UC Berkeley. JB acknowledges support from NASA grants 80NSSC18K1452 and 80NSSK0834, and NSF Grant 1602655. \mycolor{FG acknowledges supports in part from NASA grant 80HQTR20T0073, 80HQTR21T0087 and 80HQTR21T0104. YK acknowledges support from the Swedish National Space Agency.}

\bmhead{Conflict of interest}
The authors have no conflict of interest to declare that is relevant to the content of this article.

\bibliography{paper3.3.bib}



\end{document}